\documentclass[pre,twocolumn]{revtex4}

\usepackage{graphicx}
\usepackage{amssymb}
\usepackage{amsmath}
\usepackage{amsfonts}
\usepackage{mathrsfs}

\newcommand {\be} {\begin{equation}}
\newcommand {\ee} {\end{equation}}
\newcommand {\bea} {\begin{eqnarray}}
\newcommand {\eea} {\end{eqnarray}}
\newcommand {\bes} {\begin{displaymath}}
\newcommand {\ees} {\end{displaymath}}
\newcommand {\beas} {\begin{eqnarray*}}
\newcommand {\eeas} {\end{eqnarray*}}

\begin{document}

\title{Applying a potential across a biomembrane: electrostatic contribution to the bending rigidity and membrane instability}

\author{Tobias Ambj\"ornsson}
\altaffiliation[Present address: ]{Department of
  Chemistry, Massachusetts Institute of Technology, 77 Massachusetts
  Avenue, Cambridge, Massachusetts 02139.}
\author{Michael A. Lomholt}
\altaffiliation[Present address: ]{Department of Physics,
  University of Ottawa, 150 Louis Pasteur, Ottawa, Ontario K1N 6N5,
  Canada.}
\affiliation{NORDITA - Nordic Institute for
Theoretical Physics, Blegdamsvej 17, DK-2100 Copenhagen \O, Denmark.}
\thanks{These authors made an equal contribution to this work.}

\author{Per Lyngs Hansen}
\affiliation{MEMPHYS - Center for Biomembrane Physics and Department
  of Physics, University of Southern Denmark, Campusvej 55, 5230 Odense M, Denmark.}

\begin{abstract}
We investigate the effect on biomembrane mechanical properties due to
the presence an external potential for a non-conductive
non-compressible membrane surrounded by different electrolytes. By
solving the Debye-H\"uckel and Laplace equations for the electrostatic
potential and using the relevant stress-tensor we find: in (1.) {\em
the small screening length limit}, where the Debye screening length is
smaller than the distance between the electrodes, the screening
certifies that all electrostatic interactions are short-range and the
major effect of the applied potential is to decrease the membrane
tension and increase the bending rigidity; explicit expressions for
electrostatic contribution to the tension and bending rigidity are
derived as a function of the applied potential, the Debye screening
lengths and the dielectric constants of the membrane and the
solvents. For sufficiently large voltages the negative contribution to
the tension is expected to cause a membrane stretching instability. For
(2.) {\em the dielectric limit}, i.e. no salt (and small wavevectors
compared to the distance between the electrodes), when the dielectric
constant on the two sides are different the applied potential induces
an effective (unscreened) membrane charge density, whose long-range
interaction is expected to lead to a membrane undulation instability.
\end{abstract}

\maketitle 

\section{Introduction}

Biomembranes are thin fluid films composed mostly of lipids. In cells
they help separating different cellular environments and
compartments. Biomembranes are typically ``soft'', i.e., the typical
energy required to bend them is of the order the thermal energy and
membrane tension is often quite small. Softness implies that membrane
geometry can become sensitive to different perturbations, such as
alteration of the electrostatic configuration. Much effort has for
instance been devoted to calculation of the electrostatic contribution
to tension and bending rigidity for membranes with fixed charges or
surface potentials in an electrolyte solution, see \cite{andelman95}
for a review; in general the presence of fixed (and screened due to
the electrolyte) charges tend to increase bending rigidity and hence
make the membrane stiffer. However, it has also been found that when
the membrane charges are not fixed but free to rearrange themselves on
the surface and no electrolyte solution is present to screen the
interaction, a long-wavelength undulation instability can occur
\cite{Kim_Sung}, somewhat similar to DNA condensation
\cite{Golestanian}. Electric fields can also be present across
intrinsically neutral membranes. An example would be a nerve cell,
where ion pumps create a potential difference between the two sides of
the nerve cell membrane \cite{hodgkin_huxley}. Another example is
provided in laboratories by the routine formation of liposomes in a
process known as electroformation \cite{angelova86}, during which
lipid membranes are swelling from electrodes under the application of
electric fields.

In this study we investigate the electromechanical coupling of a
membrane and an applied potential. In particular, we solve the
Debye-H\"uckel and Laplace equations for the electrostatic potential
for a non-conductive, incompressible membrane between two flat
electrodes (kept at fixed potentials). On either side the membrane is
surrounded by {\em different} electrolyte solutions. From the
solutions for the potential we quantify how the corresponding induced
membrane charges change the free energy for the membrane and identify
electrostatic contributions to membrane mechanical parameters. In the
presence of an electrolyte (in the small screening length limit, see
below) we find that the electrostatic contribution to membrane bending
rigidity is positive. In the absence of added salt (the dielectric
limit) the membrane becomes unstable against long wavelength
undulations (in a somewhat similar fashion to the behaviour of the interface
between two immiscible fluids, see Ref. \cite{melcher65}) if the two
fluids surrounding the membrane have different dielectric
constants. For the symmetric dielectric case, as well as for the small
screening length limit, the membrane tension receives a negative
contribution; for a sufficiently large applied potential this
contribution would lead to a membrane stretching instability.

Several studies (see \cite{andelman95} and references therein) have
investigated the electrostatic contributions to tension and bending
rigidity for membranes having a fixed surface charge density or fixed
potential at the membrane. However, less work has been dedicated to the
effect of {\em induced} charges due to an applied potential, as
considered in the present study. The results found here complement
previous results given in \cite{lacoste06} (using coupled
hydrodynamical-electric field equations, similar to \cite{ajdari95}))
for the symmetric case of a membrane surrounded by identical
electrolytes (in the small screening length limit), to the asymmetric
case and by giving an explicit expression for the bending rigidity; in the limit of identical electrolytes on the two sides of the
membrane our expression for the tension agrees with that derived in
\cite{lacoste06}. Also, our formulation allows us to
investigate the dielectric (no salt) limit, which was not considered
in \cite{lacoste06}. However, unlike \cite{lacoste06} we do not
consider any dynamical effects. In contrast to the study \cite{sens02}
based on electrolyte conductivities, our approach through the
Debye-H\"uckel equation allow us to study the effect of a finite Debye
screening length. In Ref. \cite{sens02} a non-zero membrane conductivity
was considered, whereas we consider non-conductive membranes. Similar
to the results in \cite{sens02} we find a negative contribution to the
membrane tension in the presence of an electrolyte, although the two
results are difficult to compare because of the different mathematical
formulations.

This work is organized as follows: In
Sec. \ref{sec:general_formulation} we give the general equations
governing the electrostatic response of a non-conductive membrane of
any shape; the membrane region is described by the Laplace equation
and the electrolyte solutions on either side satisfy the
Debye-H\"uckel equations. In a standard fashion, the boundary
conditions are that the potential and the displacement fields should
be continuous. In Sec. \ref{sec:flat_membrane} these equations are
solved for the case of a flat membrane in an external potential. In
Sec. \ref{sec:curved_membrane} corrections to the flat case solutions
are derived for a weakly curved incompressible membrane. In
Sec. \ref{sec:forces} the forces acting on the membrane, as well as
the corresponding electrostatic contribution to the membrane free
energy, are obtained. Assuming that membrane fluctuations occur on a
time scale slower than the relaxation time for the electrostatic
potential we use the expressions for the forces for the weakly curved
membrane in order to obtain the renormalized membrane mechanical
parameters, such as tension and bending rigidity, (in terms of a power
series expansion in the wavevector) as a function of the applied
potential, the salt concentrations (entering through the Debye
screening lengths) and the dielectric constants of the membrane and
the solvents. We investigate three different limits: (1.) {\em the
  small screening length limit}, where the Debye screening length is
smaller than the distance between the electrodes; (2.) {\em the
  dielectric limit}, i.e., no salt; (3.) {\em the symmetric case},
where the salt concentration and dielectric constants on the two sides
of membrane are equal. The results for the membrane mechanical
parameters in the three limits above (the main results in this study)
are given in Eqs. (\ref{eq:f1_el})-(\ref{eq:Sigma_diel}). Finally, in
Sec. \ref{sec:summary_outlook} a summary and discussion are given.

\section{General formulation}\label{sec:general_formulation}

We are interested in how biomembrane mechanical parameters (and
thereby, for instance, membrane fluctuations) are effected by an
applied potential. Two parameters characterizing a membrane in the
absence of an applied potential are the tension $\sigma$ and the
bending stiffness $K$ \cite{helfrich73,seifert97}. As an external
potential is applied there will in general be electrostatic
contributions ($\sigma_{\rm el}$ and $K_{\rm el}$) to both of these
quantities so that $\sigma \rightarrow \sigma + \sigma_{\rm el}$ and
$K \rightarrow K +K_{\rm el}$ in the presence of the applied
potential. An aim of this paper is to calculate $\sigma_{\rm el}$ and
$K_{\rm el}$. In a standard fashion we consider a small perturbation
from a flat membrane, characterized by a height undulation $h(x,y)$
(where $x$ and $y$ are coordinates in the plane of the flat membrane)
and solve the electrostatic equations (via a Fourier-transformation in
the $x$- and $y$ coordinates) for this weakly perturbed
geometry. Through a power series expansion in wavevector $q$
($q=\sqrt{q_x^2+q_y^2}$, where $q_x$ and $q_y$ are the
Fourier-transform variables of $x$ and $y$ respectively) of the free
energy $G$ one may identify $\sigma_{\rm el}$ and $K_{\rm el}$ (see
chapter 2 in Ref. \cite{seifert97}). We find it convenient to, rather
than utilize the free energy directly, consider the electrostatic contribution
to the ``restoring'' force, from which we identify $\sigma_{\rm el}$
($K_{\rm el}$) as the prefactor in front of the $-q^2 \bar{h}$
($-q^4\bar{h}$) term in a small $q$-expansion of the force, where
$\bar{h}=\bar{h}(q_x,q_y)$ is the Fourier-transform of
$h(x,y)$. Notice that for obtaining the tension and bending rigidity
it suffices to keep terms linear in $\bar{h}$ in the restoring force
expression. In general there may be other terms in the power series
expansion in $q$. For instance, as noted in the introduction, for the
asymmetric dielectric case there is a membrane undulation instability which
mathematically arises due to the presence of a negative term linear in $q$ 
in the series expansion. 

The approach described above relies on a
``quasi-static'' approximation, i.e. we assume that membrane
fluctuations occur on a time scale $t_{\rm mem}$ slower than the time
scale $t_{\rm el}$ over which the electrostatic configuration adjusts
itself ($t_{\rm el}\ll t_{\rm mem}$).  This assumption allows us to
solve the electrostatic problem for a {\em fixed}, weakly curved (but
otherwise arbitrary) geometry. Let us estimate the time scales $t_{\rm
  el}$ and $t_{\rm mem}$: To estimate $t_{\rm el}$ for an electrolyte
we assume that this time scale equals the time for an ion to diffuse
the distance of the order the Debye screening length, i.e. $t_{\rm
  el}\approx \kappa^{-2}/D$, where $D$ is the ion diffusion constant
and $\kappa$ is the inverse Debye screening length introduced below
(one may more realistically assume that $t_{\rm el}=\min
\{\kappa^{-2},q^{-2}\}/D$, since for a wavelength perturbation of the
order $1/q$ the ions need only diffuse a distance $1/q$ for the ion
cloud to relax). From the Einstein relation and Stoke's law, we have
$D=k_B T/(6\pi \eta R)$, where $\eta$ is the viscosity, $k_B$ is the
Boltzmann constant, $T$ the temperature and $R$ the ion Stoke's radius, using
$k_BT=4\cdot 10^{-21}$ J, $\eta=10^{-3}$ Ns/m$^2$ and $R\approx
0.1-0.3$ nm \cite{andersson94}, we find $D\approx 10^{-9}$
m$^2$/s. Furthermore, taking $\kappa^{-1}=10$ nm we obtain: $t_{\rm
  el}\approx 10^{-7}$ s. For the membrane relaxation time we estimate
\cite{miao02} $t_{\rm mem}\approx \eta/(q^3 K)$, and assuming
$q^{-1}>100$ nm, $K=10^{-19}$ J we find $t_{\rm mem}\gtrsim 10^{-5}$
s. Therefore, indeed, we have $t_{\rm el}\ll t_{\rm mem}$ in
general. We note from the expressions above that the longer the
wavelength perturbation (smaller $q$) the better justified is our
quasi-static approximation. In the dielectric limit there are no ions
and the relevant relaxation time $t_{\rm el}$ is instead that of water
relaxation (hydrogen-bond rearrangement time), which typically is of
the order $10^{-12}$ s (see Ref. \cite{ohmine93}) at room temperature,
again certifying that $t_{\rm el}\ll t_{\rm mem}$.

\begin{figure}
\begin{center}
\includegraphics[width=8cm]{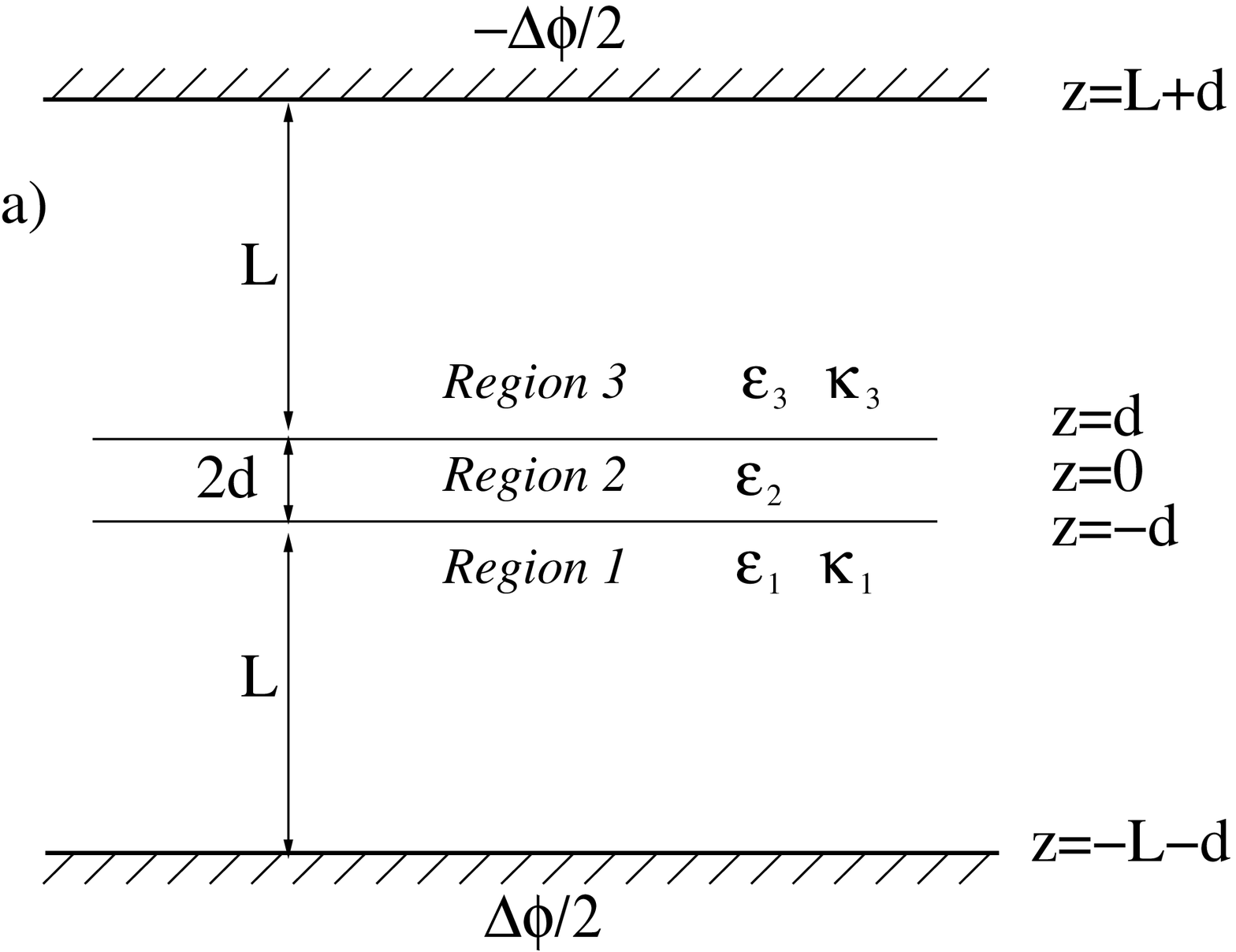}
\end{center}
\begin{center}
\includegraphics[width=8cm]{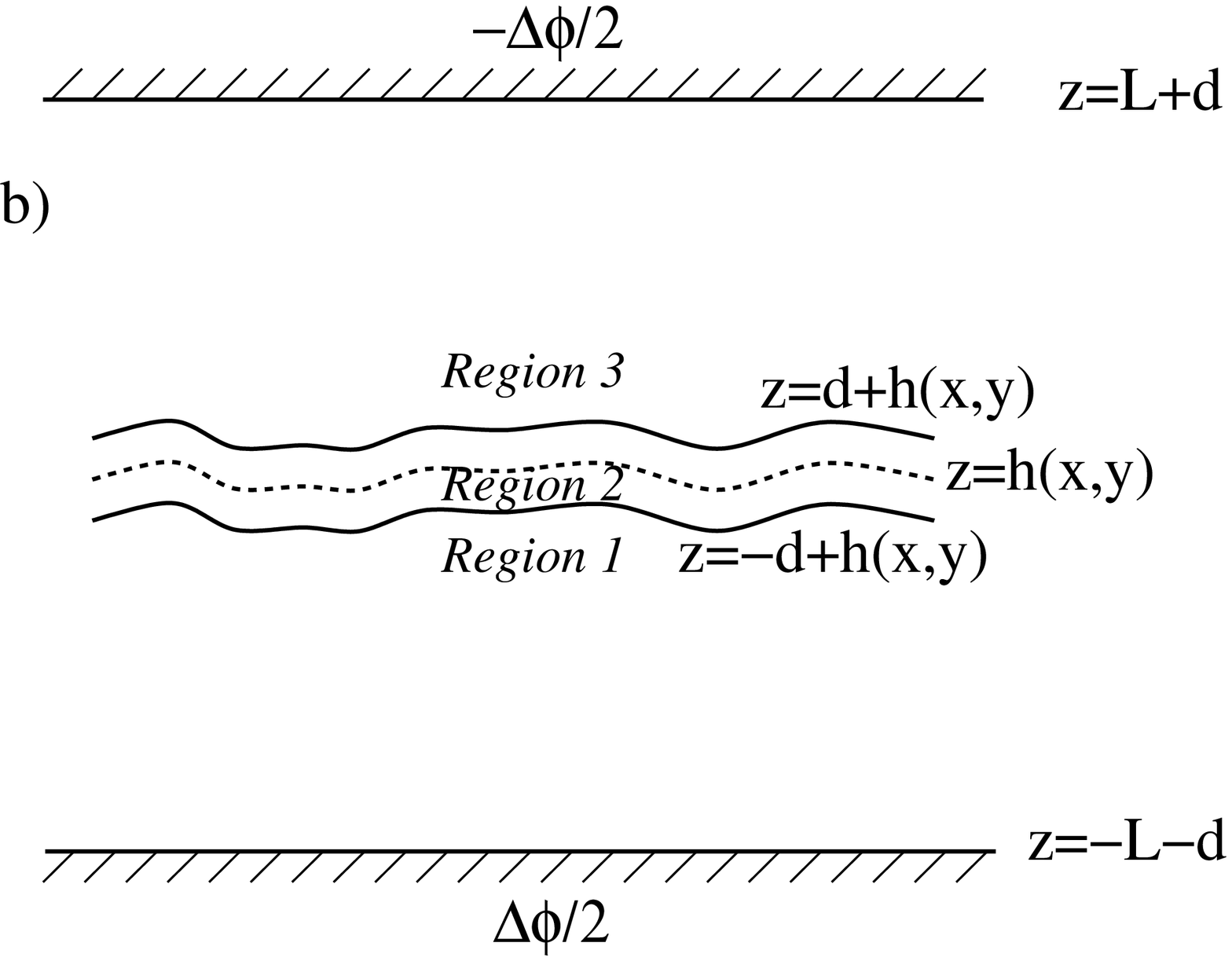}
\end{center}
\caption{Cartoon of the problems considered in this study: a membrane
of width $2d$ is placed between two electrodes. Region 1 and 3 are
characterized by dielectric constant $\varepsilon_1$ and
$\varepsilon_3$ and Debye screening lengths $\kappa_1$ and $\kappa_3$
respectively. The membrane is assumed non-conductive (and
non-compressible) and characterized by a dielectric constant
$\varepsilon_2$. Solving for the electrostatic potential
$\Phi(\vec{x})$ for (a) a flat membrane, and (b) a weakly curved
membrane (and utilizing the stress-tensor) allows us to obtain the
electrostatic contribution to the membrane mechanical parameters.}
\label{fig:cartoon}
\end{figure}

We are now set to consider the the effect of an applied potential on
the mechanical properties of a biomembrane within the quasi-static
approximation.  The explicit problem we consider is depicted in
Fig. \ref{fig:cartoon}: an incompressible membrane of thickness $2d$
is placed with its center-of-mass at positions $z=0$ between two flat
electrodes (at $z=\pm(L+d)$) which are kept at potentials $\mp\Delta
\phi/2$; the distance between the membrane surface (for a flat
membrane) and the electrodes are hence $L$. Regions 1 and 3 are
electrolyte solutions, and in general these two regions are of
different composition (different concentration of ions and different
dielectric constants). As noted above the first stage towards
calculating electrostatic forces on the membrane and thereby the
membrane free energy in the presence of the applied potential is to
obtain the electrostatic potential $\Phi(\vec{x})$. In this section we
give general equations determining $\Phi(\vec{x})$ for any membrane
shape. In the subsequent sections we analyze in detail: (i) the flat
membrane case, see Fig \ref{fig:cartoon}a); all quantities for this
case carry a superscript $(0)$; (ii) for the weakly curved situation
(since we assume the membrane to be incompressible we only consider
undulation deformation modes), illustrated in
Fig. \ref{fig:cartoon}b), there will be corrections of all quantities
compared with the flat case; all such corrections carry the
superscript $(1)$.

For all three regions the electrostatic potential 
satisfies the Poisson equation (using SI units \cite{Jackson})
  \be
\varepsilon_0 \varepsilon_\gamma \nabla^2 \Phi_\gamma (\vec{x}) = -\rho_\gamma (\vec{x})\label{eq:Poisson_eq}.
  \ee
We will henceforth use a subscript $\gamma$ (=$1$,$2$ or $3$) to
distinguish quantities in the three different regions. Above
$\varepsilon_0$ is the permittivity of vacuum, $\varepsilon_\gamma$ is
the dielectric constant for region $\gamma$, and
$\rho_\gamma(\vec{x})$ is the free charge density (the bound charges
are, in a standard fashion, taken into account through
$\varepsilon_\gamma$). We require
  \be
\int_{\rm region \ \gamma}  \rho_\gamma(\vec{x}) d^3x =0 \label{eq:charge_neutral},
  \ee
i.e, that in each region we have charge neutrality (the membrane
is assumed to be impermeable to ions).

Let us now consider the explicit expression for the charge density in
each of the three regions.\\
\noindent {\em Region 2}. In region 2 we assume
  \be
\rho_2(\vec{x})=0\label{eq:rho2},
  \ee
i.e. there are no free charges in this region.\\
\noindent {\em Region 1}. In region 1 we assume that there are ions of
two types (positively and negatively charged) which are taken to be
Boltzmann-distributed, i.e. ($i$ distinguishes the different ionic
species) $\rho_1(\vec{x})=\sum_{i} q^i c_{\rm bulk,1}^i \exp [-\beta
q^i (\Phi_1(\vec{x})-\phi^0_1)]$ where $q^i$ is the charge of ionic
species $i$, $c_{\rm bulk,1}^i$ is the concentration of ions in region
1, and $\beta=1/(k_BT)$, with $k_B$ the Boltzmann constant and $T$ the
temperature, as before. The constant $\phi^0_1$ is determined through
the charge neutrality condition. Making a linear approximation,
i.e. assuming $\beta q^i (\Phi_1(\vec{x})-\phi^0_1)\ll 1$, and using
the charge neutrality condition in the absence of the external
potential, $\sum_i q^i c_{\rm bulk}^i=0$, we find that the charge
density can be written:
  \be
\rho_1(\vec{x})=-\varepsilon_0\varepsilon_1\kappa_1^2 (\Phi_1(\vec{x})-\phi^0_1)\label{eq:rho1},
  \ee
with
  \be
\kappa_1^2=\frac{\beta}{\varepsilon_0\varepsilon_1 } \sum_{i} [q^i]^2 c_{\rm bulk,1}^i,
  \ee
where $\kappa_1$ is the inverse Debye screening length for region 1.\\
\noindent {\em Region 3}. For region 3, with the same approximations as above,
the charge density becomes:
  \be
\rho_3(\vec{x})=-\varepsilon_0\varepsilon_3 \kappa_3^2 (\Phi_3(\vec{x})-\phi^0_3)\label{eq:rho3},
  \ee
where
  \be
\kappa_3^2=\frac{\beta}{\varepsilon_0\varepsilon_3 } \sum_{i} [q^i]^2 c_{\rm bulk,3}^i
  \ee
is the square of the inverse Debye screening length and $c_{\rm
  bulk,3}^i$ is the concentration of ions in region 3.

Inserting Eqs. (\ref{eq:rho2}) into Eq. (\ref{eq:Poisson_eq}) yields the
Laplace equation
  \be
\nabla^2 \Phi_2(\vec{x})=0\label{eq:Laplace}
  \ee
for the potential in region 2. Inserting (\ref{eq:rho1}) and
(\ref{eq:rho2}) into Eq. (\ref{eq:Poisson_eq}) gives the Debye-H\"uckel
equation ($\gamma=1,3$)
  \be
\nabla^2 \Phi_\gamma(\vec{x})-\kappa_\gamma^2 (\Phi_\gamma (\vec{x})-\phi^0_\gamma)=0\label{eq:DH}
  \ee
for regions 1 and 3. The charge neutrality condition,
Eq. (\ref{eq:charge_neutral}), for region 1 and 3 becomes
  \be
\int_{\rm region \ \gamma} [\Phi_\gamma (\vec{x})-\phi^0_\gamma] d^3 x=0\label{eq:charge_neutral2}.
  \ee
Note that we have trivial charge neutrality in region 2 [see Eq. (\ref{eq:rho2})].

Let us now consider the boundary conditions supplementing the
equations above. At the electrodes we have:
  \bea
\Phi_1(\vec{x})|_{z=-L-d}&=&\frac{\Delta \phi}{2},\nonumber\\
\Phi_3(\vec{x})|_{z=L+d}&=&-\frac{\Delta \phi}{2}.\label{eq:BC1}
  \eea
In addition we have that the potential and the normal component of the
displacement fields are continuous across the region 1-region 2 and
region 2-region 3 boundaries, i.e. \cite{Jackson}
  \bea
\Phi_1(\vec{x})&=&\Phi_2(\vec{x})\ {\rm at \ region \ 1-2 \ boundary},\nonumber\\
\Phi_2(\vec{x})&=&\Phi_3(\vec{x})\ {\rm at \ region \ 2-3 \ boundary}\label{eq:BC2}
  \eea
and
 \bea
\varepsilon_1 \hat{n}\cdot \vec{\nabla}\Phi_1(\vec{x})&=&\varepsilon_2 \hat{n}\cdot \vec{\nabla}\Phi_2(\vec{x})\ {\rm at \ region \ 1-2 \ boundary},\nonumber\\
\varepsilon_2 \hat{n}\cdot \vec{\nabla}\Phi_2(\vec{x})&=&\varepsilon_3 \hat{n}\cdot \vec{\nabla}\Phi_3(\vec{x})\ {\rm at \ region \ 2-3 \ boundary},\nonumber\\ \label{eq:BC3}
  \eea
where $\hat{n}$ is the normal to the respective
interface. Eqs. (\ref{eq:Laplace}), (\ref{eq:DH}) and
(\ref{eq:charge_neutral2}), together with the boundary conditions
Eqs. (\ref{eq:BC1}), (\ref{eq:BC2}) and (\ref{eq:BC3}) completely
determine the electrostatic potential $\Phi(\vec{x})$.  

From the solutions for $\Phi(\vec{x})$ one can calculate other
quantities. For instance, one may obtain the {\rm induced} potential
defined through $\Phi_{\rm ind}(\vec{x})=\Phi(\vec{x})-\Phi_{\rm
appl}(\vec{x})$ , where the applied potential is $\Phi_{\rm
appl}(\vec{x})=-\Delta \phi z/[2(L+d)]$. The total
electric field is given by
$\vec{E}(\vec{x})=-\vec{\nabla}\Phi(\vec{x})$, the applied electric
field is $\vec{E}_{\rm appl}=-\vec{\nabla}\Phi_{\rm appl}$ and the
induced field is $\vec{E}_{\rm ind}=-\vec{\nabla}\Phi_{\rm
ind}=\vec{E}-\vec{E}_{\rm appl}$. In the next section we solve the
equations given in this section for the case of a flat membrane. In
the section after we find corrections to the flat case solutions for a
weakly curved membrane.

\section{Potential for a flat membrane}\label{sec:flat_membrane}

Below we obtain the electrostatic potential in and around a flat
membrane. We use a superscript $(0)$ to indicate the flat case
quantities. 

For a flat membrane the solutions depend only on $z$, and
explicitly the solutions to Eqs. (\ref{eq:Laplace}) and (\ref{eq:DH})
are
  \be
\Phi_2^{(0)}(z)=\phi^0_2+A_2 z\label{eq:Phi_flat_2}
  \ee
where $\phi^0_2$ and $A_2$ are constants determined by the boundary conditions below. Also,
  \be
\Phi_1^{(0)}(z)=\phi^0_1+A_1(e^{\kappa_1(z+d)}-e^{-\kappa_1(L+d+z)})\label{eq:Phi_flat_1}
  \ee
and 
  \be
\Phi_3^{(0)}(z)=\phi^0_3+A_3(e^{-\kappa_3(z-d)}-e^{-\kappa_3(L+d-z)})\label{eq:Phi_flat_3}
  \ee
where $\phi^0_1$, $\phi^0_3$, $A_1$ and $A_3$ are constants, and we used the charge
neutrality condition, Eq. (\ref{eq:charge_neutral2}).

We now use the boundary conditions (together with the fact that the
boundary surfaces are at $z=\pm d$ for the flat case considered here,
see Fig \ref{fig:cartoon}a) in order to determine the unknown
constants above. From Eq. (\ref{eq:BC1}), the condition that the
potential is continuous, Eq. (\ref{eq:BC2}) and the fact that the
displacement field is continuous, Eq. (\ref{eq:BC3}), we get 6
equations for the 6 constants $\phi_\gamma^0$ and $A_\gamma$
($\gamma=1,2,3$). Solving these equations leads to
   \bea
   A_1&=&-\frac{\Delta \phi}{2(1+e^{-\kappa_1 L})} l_1 \Gamma \nonumber\\
   A_2&=&-\frac{\Delta \phi}{2 \varepsilon_2} \Gamma \nonumber\\
   A_3&=& \frac{\Delta \phi}{2(1+e^{-\kappa_3 L})} l_3 \Gamma \label{eq:coeff_flat_1}
  \eea
and 
  \bea
\phi^0_1&=&\frac{\Delta \phi}{2} \Big( 1- g(\kappa_1)  l_1  \Gamma \Big) \nonumber\\
\phi^0_2&=&-\frac{\Delta \phi}{2} \Big(  g(\kappa_1) l_1 - g(\kappa_3)  l_3 \Big) \Gamma \nonumber\\
\phi^0_3&=&-\frac{\Delta \phi}{2} \Big( 1- g(\kappa_3)  l_3  \Gamma\Big) \label{eq:coeff_flat_2}
  \eea
where 
 \be
g(q)=\frac{1-e^{-qL}}{1+e^{-qL}}=\tanh(\frac{qL}{2})\label{eq:g_q}
  \ee
and $\Gamma= 1/[ g(\kappa_1) l_1 + g(\kappa_3)l_3 +d/\varepsilon_2] $,
and we introduced the ``rescaled" Debye screening lengths
$l_1=(\varepsilon_1\kappa_1)^{-1}$ and
$l_3=(\varepsilon_3\kappa_3)^{-1}$.

There are three limits of particular interest: 
\begin{enumerate}
\item  {\em ``small'' screening
length}, $\kappa_1L, \kappa_3L \gg 1$. For this case we have
$g(q)\rightarrow 1$ in Eqs. (\ref{eq:coeff_flat_1}) and
(\ref{eq:coeff_flat_2}). Also the prefactors for $A_1$ and $A_3$
simplify.
\item we define {\em the dielectric limit} as the limit of no salt,
i.e. $c_{\rm bulk,1}^i, c_{\rm bulk,3}^i \rightarrow 0$; within the
Debye-H\"uckel approximation this is the equivalent to
$\kappa_1,\kappa_3\rightarrow 0$. Expanding the
exponentials in Eqs. (\ref{eq:Phi_flat_1}) and
(\ref{eq:Phi_flat_3}), and using the explicit form for $A_1$ and
$A_3$ above one straightforwardly show that the solutions in all three
regions take the form $\Phi(\vec{x})=az+b$ where $a$ and $b$ are
constants independent of $\kappa_1$ and $\kappa_3$, as it should since
in the dielectric limit the potential satisfies Laplace equation,
$\nabla^2 \Phi (\vec{x})=0$, see Eq. (\ref{eq:DH}).
\item {\em the symmetric case},
$\varepsilon_1=\varepsilon_3$ and $\kappa_1=\kappa_3$. In this limit
we find that $A_1=-A_3$, $\phi^0_1=-\phi^0_3$ and $\phi^0_2=0$.
\end{enumerate}

\begin{figure}
\begin{center}
\includegraphics[width=6.2cm]{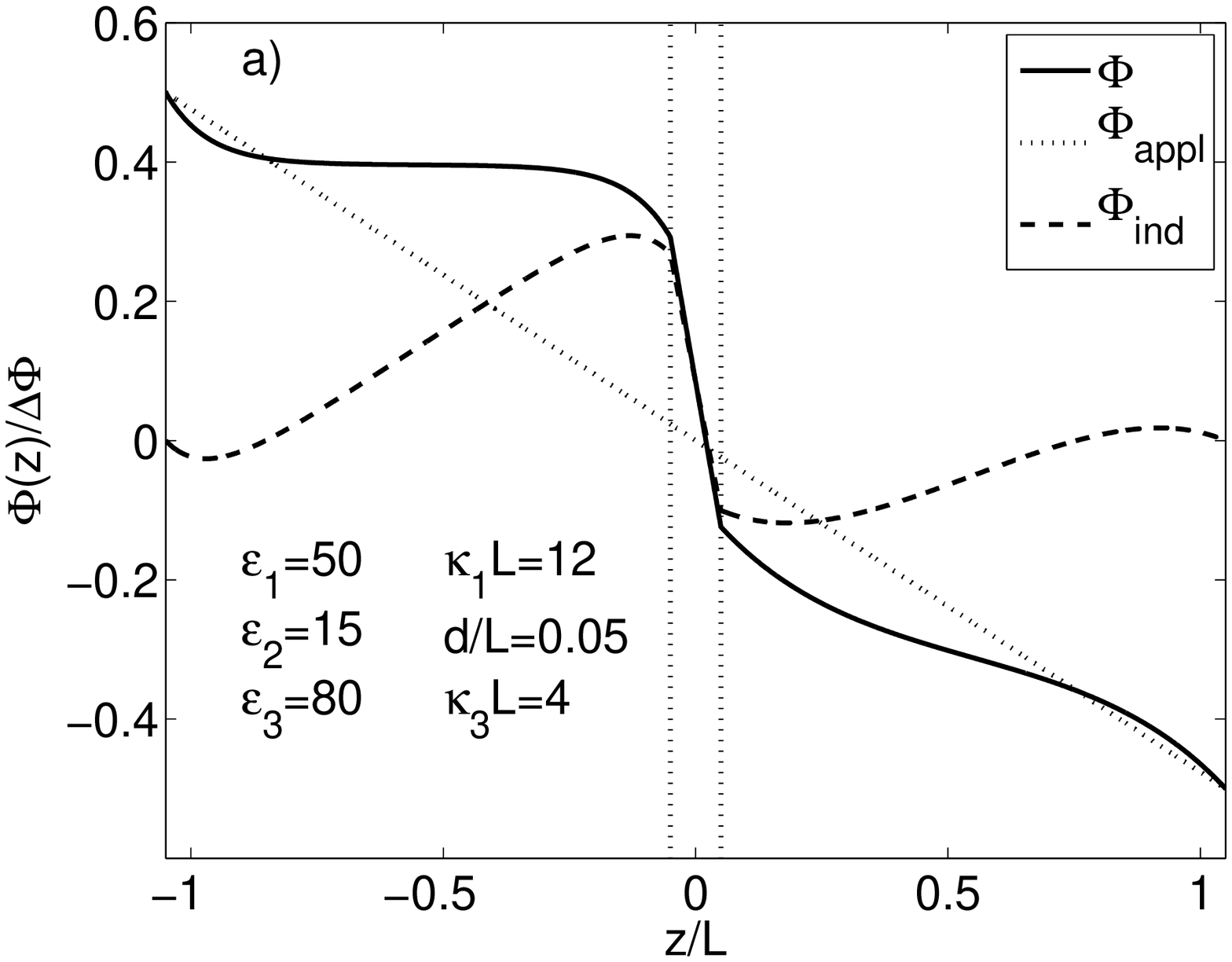}
\end{center}
\begin{center}
\includegraphics[width=6.2cm]{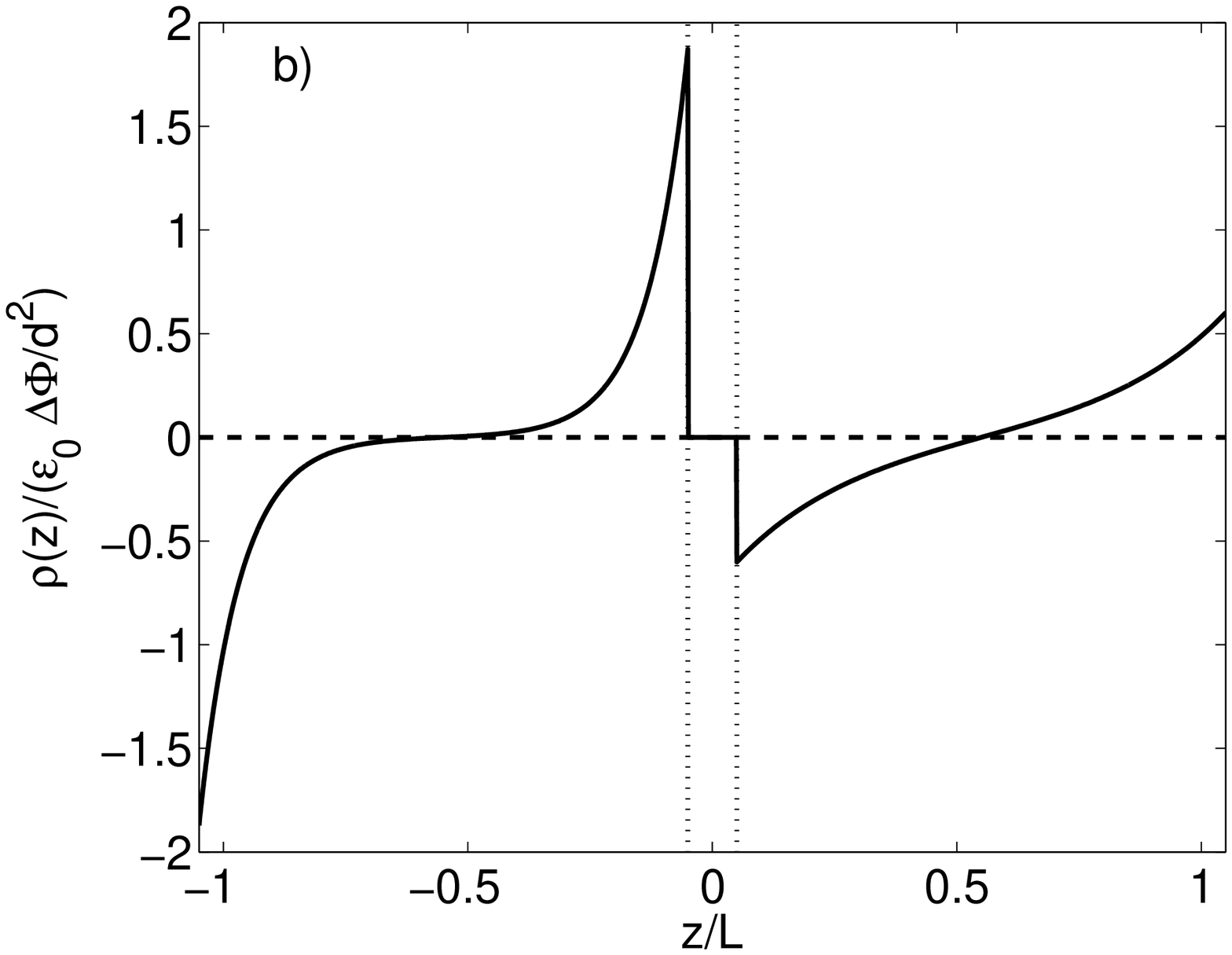}
\end{center}
\begin{center}
\includegraphics[width=6.2cm]{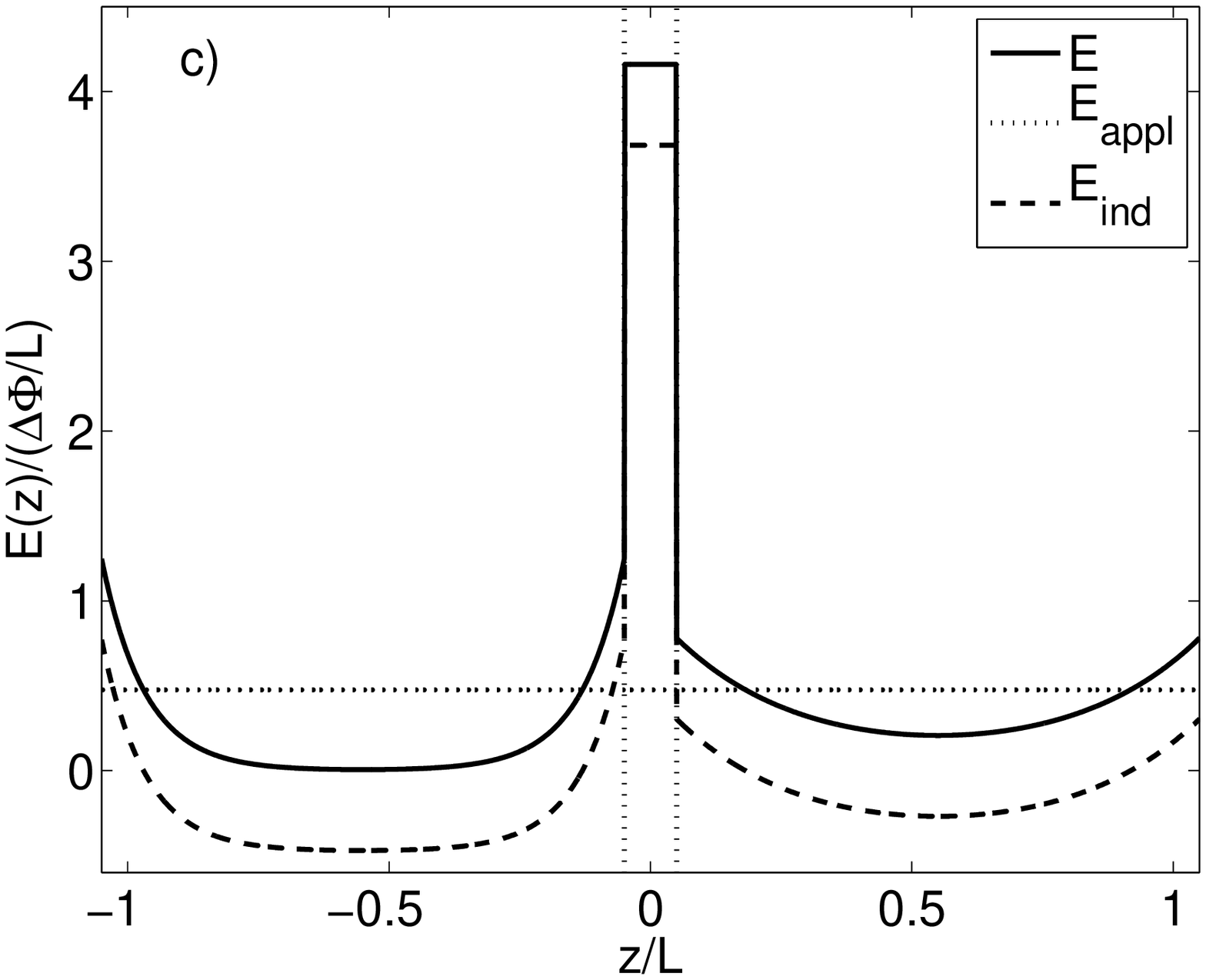}
\end{center}
\caption{Illustration of the response of a flat membrane,
surrounded by different electrolyte solutions, to an applied
potential. a) Electrostatic potential $\Phi(\vec{x})=\Phi^{(0)}(z)$
for a flat membrane (solid curve) as a function of position
$z$. Illustrated are also the applied potential $\Phi_{\rm appl}(z)$
(dotted curve) and the induced potential $\Phi_{\rm ind}(z)$ (dashed
curve); b) Free charge density $\rho(z)$ (solid curve) as a function
of position $z$. The dashed curve at $\rho\equiv 0$ corresponds to the
free charge density in the absence of the applied potential ($\Delta
\phi=0$); c) The electric field $E=-\partial \Phi/\partial z$ in the
$z$-direction for different positions $z$. Illustrated are also the
applied field $E_{\rm appl}=-\partial\Phi_{\rm appl}/\partial z$ and
the induced field $E_{\rm ind}=-\partial \Phi_{\rm ind}/\partial
z$. The vertical lines in each figure corresponds to the membrane
surfaces. The chosen parameter values are the same for all three
graphs, and are listed in the top graph (we chose the somewhat
unrealistically large value $\varepsilon_2=15$ in order to be able to
better visualize the Debye screening layers, a more reasonable value
would be $\varepsilon_2\approx 2$). Note that the potential is
continuous, whereas the free charge density and the $z$-component of
the electric field are not continuous, as it should.}
\label{fig:pot}
\end{figure}

The potential and charge densities are illustrated in
Fig. \ref{fig:pot}, using the flat membrane results in
Eqs. (\ref{eq:Phi_flat_2}), (\ref{eq:Phi_flat_1}) and
(\ref{eq:Phi_flat_3}). The electric field in the $z$-direction (the
electric field components in the $x$- and $y$-direction are zero for a
flat membrane) is also illustrated.  We notice that the potential is
continuous as it should and that the free charges tend to build up
close to the membrane and electrodes (for $\kappa_1,\kappa_3\neq
0$). Since the normal component of the displacement field is
continuous across the boundaries, the relative jump in the electric
field as the the boundary between region 1-2 (region 2-3) is crossed
equals $\varepsilon_2/\varepsilon_1$ ($\varepsilon_2/\varepsilon_3$),
see Fig. \ref{fig:pot} (bottom).

\section{Potential for a weakly curved membrane}\label{sec:curved_membrane}

We now consider a weakly curved membrane, see Fig \ref{fig:cartoon}b):
the center of the membrane is slightly displaced from the flat ($z=0$)
case, according to $z=h(x,y)$ with membrane surfaces at $z=\pm d+
h(x,y)$ \cite{note}. We write the solution for the electrostatic
potential according to ($\gamma=1,2,3$ as before):
  \be
\Phi_\gamma(\vec{x})=\Phi_\gamma^{(0)}(z)+\Phi_\gamma^{(1)}(\vec{x})
\label{eq:full_solution},
  \ee
where $\Phi_\gamma^{(0)}(z)$ is the potential for region $\gamma$ for
the flat case given in the previous section and
$\Phi_\gamma^{(1)}(\vec{x})$ is a correction to the potential due to
the perturbed geometry. In this section we will calculate
$\Phi_\gamma^{(1)}$ only to first order in the perturbation $h(x,y)$;
this suffices for obtaining membrane mechanical parameters such as
tension and bending rigidity (see discussion at the begininning of
Sec. \ref{sec:general_formulation}). In each of the regions
Eqs. (\ref{eq:Laplace}) and (\ref{eq:DH}) has to be satisfied.
$\Phi_\gamma^{(0)}(\vec{x})$ satisfies these equations for
$h(x,y)\equiv 0$, and we therefore require that the corrections
satisfy:
  \be
\nabla^2 \Phi_2^{(1)}(\vec{x})=0\label{eq:Laplace_curved}
  \ee
and ($\gamma=1,3$)
  \be
\nabla^2 \Phi_\gamma^{(1)}(\vec{x})-\kappa_\gamma^2 \Phi_\gamma^{(1)}(\vec{x})=0\label{eq:DH_curved}.
  \ee
Let us now consider the boundary conditions (the boundary surfaces are
at $z=\pm d+h(x,y)$ for the curved membrane considered here). Since
$\Phi_\gamma^{(0)}(z)$ satisfies Eq. (\ref{eq:BC1}) we require
for the perturbation:
  \bea
\Phi_1^{(1)}(\vec{x})|_{z=-L-d}&=&0,\nonumber\\
\Phi_3^{(1)}(\vec{x})|_{z=L+d}&=&0.\label{eq:BC1_curved}
  \eea
Before imposing the conditions that the potential and the displacement
fields are continuous, Eqs. (\ref{eq:BC2}) and (\ref{eq:BC3}), we note
that for ``small'' $h$ any scalar quantity $g(\vec{x})$ may be
expanded, to first order in $h$, according to:
  \beas
g(\vec{x})&=&g^{(0)}(\vec{x})|_{z=\pm d+h}+g^{(1)}(\vec{x})|_{z=\pm d+h}\nonumber\\
&\approx & \big( g^{(0)}(\vec{x})+h \frac{\partial g^{(0)}(\vec{x})}{\partial z}+g^{(1)}(\vec{x})\big) |_{z=\pm d},
  \eeas
we briefly discuss the quantitative meaning of ``small'' $h$ at the
end of this section. Eqs. (\ref{eq:BC2}) and (\ref{eq:BC3}) can then
be written
  \bea
\big( h\frac{\partial \Phi_1^{(0)}}{\partial z}+\Phi_1^{(1)}\big)|_{z=-d}&=&\big( h\frac{\partial \Phi_2^{(0)}}{\partial z}+\Phi_2^{(1)} \big)|_{z=-d},\nonumber\\
\big( h\frac{\partial \Phi_2^{(0)}}{\partial z}+\Phi_2^{(1)} \big)|_{z=d}&=&\big( h\frac{\partial \Phi_3^{(0)}}{\partial z}+\Phi_3^{(1)}\big)|_{z=d} \label{eq:BC2_curved}
  \eea
and 
  \bea
\varepsilon_1 \big( h\frac{\partial^2 \Phi_1^{(0)}}{\partial z^2}+\frac{\partial \Phi_1^{(1)}}{\partial z} \big)|_{z=-d}&=&\varepsilon_2 \big( h\frac{\partial^2 \Phi_2^{(0)}}{\partial z^2}+\frac{\partial \Phi_2^{(1)}}{\partial z} \big)|_{z=-d},\nonumber\\
\varepsilon_2 \big( h\frac{\partial^2 \Phi_2^{(0)}}{\partial z^2}+\frac{\partial \Phi_2^{(1)}}{\partial z} \big)|_{z=d}&=&\varepsilon_3 \big( h\frac{\partial^2 \Phi_3^{(0)}}{\partial z^2}+\frac{\partial \Phi_3^{(1)}}{\partial z} \big)|_{z=d},\nonumber\\ 
\label{eq:BC3_curved}
  \eea
where we used the fact that $\Phi_\gamma^{(0)}$ satisfies the boundary
conditions for $h=0$. Eqs. (\ref{eq:Laplace_curved}) and
(\ref{eq:DH_curved}) together with the boundary conditions
Eqs. (\ref{eq:BC1_curved}), (\ref{eq:BC2_curved}) and
(\ref{eq:BC3_curved}) completely determine the correction
$\Phi_\gamma^{(1)}(\vec{x})$.

We proceed by introducing the Fourier-transform in the $x$- and
$y$-direction (not for $z$-direction) of $\Phi_\gamma^{(1)}(\vec{x})$:
  \be
\bar{\Phi}_\gamma^{(1)}(q_x,q_y,z)=\int dx dy\; e^{iq_x x+iq_y y} \Phi_\gamma^{(1)}(\vec{x}),
  \ee
and similarly we denote by ${\bar h}(q_x,q_y)$ the Fourier-transform of $h(x,y)$.
Eqs. (\ref{eq:Laplace_curved}) and (\ref{eq:DH_curved}) can then
be written:
  \be
\frac{\partial \bar{\Phi}_2^{(1)}}{\partial z^2}-q^2 \bar{\Phi}_2^{(1)}=0\label{eq:Laplace_Fourier}
  \ee
and ($\gamma=1,3$)
  \be
\frac{\partial \bar{\Phi}_\gamma^{(1)}}{\partial z^2}-\bar{q}_\gamma^2 \bar{\Phi}_2^{(1)}=0\label{eq:DH_Fourier},
  \ee
where $q=\sqrt{q_x^2+q_y^2}$ and 
  \be
\bar{q}_\gamma=\sqrt{q^2+\kappa_\gamma^2}.
  \ee 
The boundary conditions Eqs. (\ref{eq:BC1_curved}),
(\ref{eq:BC2_curved}) and (\ref{eq:BC3_curved}) remains the same in
Fourier-space (using the fact that $\Phi_\gamma^{(0)}$ is independent
of $x$ and $y$), with the sole replacement
$\Phi_\gamma^{(1)}\rightarrow \bar{\Phi}_\gamma^{(1)}$ and
$h\rightarrow {\bar h}$. The solution of
Eq. (\ref{eq:Laplace_Fourier}) is
  \be
\bar{\Phi}_2^{(1)}(q,z)=C_2(q) e^{qz}+D_2(q)e^{-qz}\label{eq:bar_Phi2},
  \ee
with $q$-dependent coefficients $C_2(q)$ and $D_2(q)$.  Using the
boundary condition Eq. (\ref{eq:BC1_curved}) we find that the solutions
to Eq. (\ref{eq:DH_Fourier}) are:
  \be
\bar{\Phi}_1^{(1)}(q,z)=D_1(q)\big(e^{\bar{q}_1(z+d)}-e^{-\bar{q}_1(z+d+2L)}\big)\label{eq:bar_Phi1}
  \ee
and 
   \be
\bar{\Phi}_3^{(1)}(q,z)=D_3(q)\big( e^{-\bar{q}_3(z-d)}-e^{\bar{q}_3(z-d-2L)}\big)\label{eq:bar_Phi3}.
  \ee
The unknown coefficients $C_2(q)$, $D_1(q)$, $D_2(q)$ and $D_3(q)$ are
determined through the boundary conditions Eqs. (\ref{eq:BC2_curved})
and (\ref{eq:BC3_curved}). We find
  \bea
D_2(q)&=&-{\bar h}\nonumber\\
& &\hspace{-1.2cm}\times \frac{e^{qd}k_3^+(p_1+\varepsilon_1 \bar{q}_1 r(\bar{q}_1)m_1^-)-e^{-qd}k_1^-(p_3-\varepsilon_3 \bar{q}_3 r(\bar{q}_3)m_3^+)}{e^{2qd}k_1^+k_3^+ - e^{-2qd} k_1^- k_3^-},\nonumber\\
\label{eq:coeff_curved_1}
  \eea
where we introduced the short-hand notations ($\gamma=1,3$)
  \bea
k_\gamma^\pm&=&k_\gamma^\pm(q)=\varepsilon_2q \pm \varepsilon_\gamma \bar{q}_\gamma r(\bar{q}_\gamma),\nonumber\\
m_\gamma^\pm&=& A_2 \pm A_\gamma \kappa_\gamma (1+e^{-\kappa_\gamma L}),\nonumber\\
p_\gamma&=& \varepsilon_\gamma A_\gamma \kappa_\gamma^2 (1-e^{-\kappa_\gamma L}),\nonumber\\
r(q)&=&\frac{1+e^{-2qL}}{1-e^{-2qL}},\label{eq:k_etc}
  \eea
with $A_\gamma$ given in Eq. (\ref{eq:coeff_flat_1}).
The remaining coefficients are
  \bea
C_2(q)&=&\frac{e^{qd}}{k_1^-} \big( D_2(q) e^{qd} k_1^+ + {\bar h}p_1 + {\bar h}\varepsilon_1 \bar{q}_1 r(\bar{q}_1) m_1^- \big),\nonumber\\
D_1(q)&=&\frac{1}{1-e^{-2\bar{q}_1 L}}\big( {\bar h}m_1^- +C_2(q) e^{-qd}+ D_2(q) e^{qd} \big), \nonumber\\
D_3(q)&=&\frac{1}{1-e^{-2\bar{q}_3 L}}\big( {\bar h}m_3^+ +C_2(q) e^{qd}+ D_2(q) e^{-qd} \big).\nonumber\\
 \label{eq:coeff_curved_2}
  \eea
We notice that $C_1(q)$, $D_1(q)$, $D_2(q)$ and $D_3(q)$ are all
proportional to ${\bar h}$ as they should be. The full solution for the
electrostatic potentials $\Phi_\gamma (\vec{x})$ for a weakly curved membrane
is given by Eq. (\ref{eq:full_solution}), where $\Phi_\gamma^{(0)}(z)$
were given in the previous section and the Fourier-transform of
$\Phi_\gamma^{(1)}(\vec{x})$ are given by Eqs. (\ref{eq:bar_Phi2}),
(\ref{eq:bar_Phi1}) and (\ref{eq:bar_Phi3}) together with
Eqs. (\ref{eq:coeff_curved_1}), (\ref{eq:k_etc}) and
(\ref{eq:coeff_curved_2}).

Again, we point out that in order to obtain membrane mechanical
parameters, such as tension and bending rigidity, it suffices to know
the restoring force (calculated in the next section) to first order in
$\bar{h}$, i.e. it is enough to consider ``small'' fluctuation
amplitudes. Therefore, even thought the results given above formally
assume that $h$ is smaller than all other length scales in the problem
($d$, $\kappa_\gamma^{-1}$, $q^{-1}$ and $L$), they are sufficient for
the purpose of calculating the tension and bending rigidity.  A
different matter is whether the Helfrich form of the electrostatic
contribution to the restoring force (in terms tension and bending
rigidity), derived in the next section, describe well ``large''
membrane fluctuations in the presence of an applied potential. To
address this question we must clarify the meaning of ``small'' $h$,
i.e. make clear what is the relevant dimensionless expansion
parameter, in the context of calculating the electrostatic
contribution to the membrane restoring force or membrane free
energy. The electrostatic contribution to the membrane free energy is
determined by the interactions between induced charges at or in the
vicinity of the membrane, with characteristic interaction distances of
the order $d$ and $\kappa_\gamma^{-1}$ (here we consider the small
screening length limit, where $\kappa_\gamma$ is non-zero).
Therefore, whenever $d q\ll 1$, $\kappa_\gamma^{-1} q \ll 1$ and $h
q\ll 1$ all interactions are effectively local on a locally flat
membrane and the free energy must take the Helfrich form
\cite{helfrich73}. Note that this argument is valid independent of
particular values of $h \kappa_\gamma$ and $h/d$, and the free energy
expansions given in the next section (in the small screening length
limit) is therefore an expansion in the (small) parameters $h q$, $d
q$, $\kappa_\gamma^{-1} q$, but when these parameters are small there
is no restriction on the values of $h \kappa_\gamma$ and $h/d$. In
consistency with the discussion above, we point out that in
Ref. \cite{goldstein90} a method that formally avoids the assumptions
$h \kappa_\gamma\ll 1$ and $h/ d\ll 1$, by utilizing the geometrical
transformation $z'=z-h(x,y)$, was found to give results consistent
with results obtained by the flat membrane perturbative approach (of
the kind used in this paper) for a charged membrane in an electrolyte.

\section{Forces, free energy and electrostatic contribution to
  membrane mechanical parameters}\label{sec:forces}

In this section we derive general expressions for the membrane forces
(within the quasi-static approximation) and the corresponding
electrostatic contribution to the membrane free energy, using the
results from the previous two sections. In particular, we obtain the
electrostatic contribution to the membrane free energy for
three cases: (1.) the {\em small screening length limit}, where the Debye
screening length is smaller than the distance between the electrodes; (2.)
{\em the dielectric limit}, i.e., no salt, and; (3.) {\em the symmetric
case}, where the salt concentrations and dielectric constants on the
two sides of the membrane are equal.

\subsection{Membrane forces via a stress-tensor calculation}

The forces acting on the membrane are obtained using the relevant
stress-tensor, $T_{ij}$. Following the derivation in Appendix
\ref{sec:derivations} we have (in the Debye-H\"uckel regime considered
here)
\bea
T_{ij}&=&\varepsilon_0 \varepsilon_\gamma \bigg( \frac{\partial \Phi_\gamma}{\partial x_i}\frac{\partial \Phi_\gamma}{\partial x_j}-\frac{1}{2} \delta_{ij} \sum_k  \frac{\partial \Phi_\gamma}{\partial x_k}\frac{\partial \Phi_\gamma}{\partial x_k}\nonumber\\
&&-\delta_{ij} \frac{\kappa_\gamma^2}{2} \big( \Phi_\gamma -\phi^0_\gamma \big)^2\bigg) -p^0_\gamma \delta_{ij}\;,\label{eq:sigmaij}
  \eea
 where $x_1=x$, $x_2=y$ and $x_3=z$. The first two terms are the usual
 Maxwell stress-tensor \cite{Jackson,landau}, the third term is an
 osmotic contribution for the ions being ``confined" by the electric
 potential (see appendix \ref{sec:derivations}), and the last term
 incorporates pressures $p_\gamma^0$ for each of the three regions
 ($\gamma=1,2,3$). The discontinuities of $T_{ij}$ at the region
 boundaries will produce forces on the membrane which will have to be
 balanced by other forces in the system, such as for example viscous
 forces within the membrane or from the surrounding bulk fluids. We
 are only interested in calculating the electromechanical contribution
 at a given $(x,y)$ to this total force balance here. To do this we
 note that the force (per unit area) in the $i$-direction on a region
 boundary from the stress in a given region is $\pm\sum_j n_jT_{ji}$
 evaluated at the boundary, where the membrane normal $n_j$ is taken
 to point towards positive $z$ and the plus (minus) sign applies when
 the region is at larger (smaller) $z$ than the boundary. Defining
  \bea
f_\gamma&=&f_\gamma^{(0)}+f_\gamma^{(1)}\nonumber\\
f_\gamma^{(0)}&=&\frac{1}{2}\varepsilon_0 \varepsilon_\gamma  \big( (\frac{\partial \Phi_\gamma^{(0)}}{\partial z})^2 - \kappa_\gamma^2  (\Phi_\gamma^{(0)} -\phi^0_\gamma )^2 \big)|_{z=\pm d}\nonumber\\
&&-p^0_\gamma\nonumber\\
f_\gamma^{(1)}&=& \varepsilon_0 \varepsilon_\gamma  \big(  \frac{\partial \Phi_\gamma^{(0)}}{\partial z} [h \frac{\partial^2 \Phi_\gamma^{(0)}}{\partial z^2} + \frac{\partial \Phi_\gamma^{(1)}}{\partial z}]\nonumber\\
& &- \kappa_\gamma^2 (\Phi_\gamma^{(0)} -\phi^0_\gamma )  [h \frac{\partial \Phi_\gamma^{(0)}}{\partial z} + \Phi_\gamma^{(1)} ] \big)|_{z=\pm d},\nonumber\\
  \eea
where $z=-d$ ($z=d$) is to be used for forces acting on the interface
separating region 1 and 2 (region 2 and 3), and using that to first
order in $h$: $n_z=1$, $n_j=- \partial_j h$ ($j=x,y$), we find that
the $z$-component of the total force acting on the surface separating
regions 1 and 2 is $f_{1-2}=f_2-f_1$. Using the explicit expressions
for the potentials from the previous two sections we find
\bea
f_{1-2}&=&f_{1-2}^{(0)}+f_{1-2}^{(1)}\nonumber\\
f_{1-2}^{(0)}&=&-\varepsilon_0 \big( 2\varepsilon_1 A_1^2\kappa_1^2 e^{-\kappa_1L}-\frac{\varepsilon_2 A_2^2}{2} \big)+p^0_1-p^0_2 \nonumber\\
\bar{f}_{1-2}^{(1)}&=&-\varepsilon_0 \big(D_1(q) \varepsilon_1 A_1 \kappa_1 [\bar{q}_1(1+e^{-\kappa_1L})(1+e^{-2\bar{q}_1 L})\nonumber\\
& & \hspace{2cm}-\kappa_1 (1-e^{-\kappa_1L})(1-e^{-2\bar{q}_1 L})]\nonumber\\
& &-\varepsilon_2 A_2 q [C_2(q)e^{-qd}-D_2(q)e^{qd}] \big)\label{eq:f12},
  \eea
where $f_{1-2}^{(0)}$ is the force on a flat membrane interface, and
$f_{1-2}^{(1)}$ is the first order correction for a weakly curved
membrane, here expressed explicitly in terms of its
Fourier-transform $\bar{f}_{1-2}^{(1)}$; note that
$\bar{f}_{1-2}^{(1)}$ is proportional to ${\bar h}$ (via $D_1$, $C_2$
and $D_2$), as it should. Similarly, the $z$-component of the force
acting on the surface separating regions 2 and 3 is $f_{2-3}=f_3-f_2$
and we find
  \bea
f_{2-3}&=&f_{2-3}^{(0)}+f_{2-3}^{(1)}\nonumber\\
f_{2-3}^{(0)}&=&\varepsilon_0 \big( 2\varepsilon_3 A_3^2\kappa_3^2 e^{-\kappa_3L}-\frac{\varepsilon_2 A_2^2}{2} \big) +p^0_2-p^0_3\nonumber\\
\bar{f}_{2-3}^{(1)}&=&\varepsilon_0 \big(D_3(q) \varepsilon_3 A_3 \kappa_3 [\bar{q}_3(1+e^{-\kappa_3L})(1+e^{-2\bar{q}_3 L})\nonumber\\
& & \hspace{2cm}-\kappa_3 (1-e^{-\kappa_3L})(1-e^{-2\bar{q}_3 L})]\nonumber\\
& &-\varepsilon_2 A_2 q [C_2(q)e^{qd}-D_2(q)e^{-qd}] \big)\label{eq:f23}
  \eea
where $A_\gamma$ are given in Eq. (\ref{eq:coeff_flat_1}) and
$C_2(q)$, $D_\gamma(q)$ are given in Eqs. (\ref{eq:coeff_curved_1})
and (\ref{eq:coeff_curved_2}). The results given in Eqs (\ref{eq:f12})
and (\ref{eq:f23}) completes the calculation of the forces acting on
the membrane interfaces. In appendix \ref{sec:force_flat_membrane} we
utilize these results in order obtain results for the total net force
on a flat membrane in some detail.  In the next subsection the
membrane free energy and the electrostatic contribution to the
membrane mechanical parameters in different limits are investigated.

\subsection{Contribution to the free energy of the membrane}\label{sec:free_energy}

Let us now investigate the electrostatic contribution to the membrane
free energy. We first note that if the fluids surrounding the membrane
are incompressible, then the pressures $p^0_\gamma$ [occurring in the
zero order terms in Eqs. (\ref{eq:f12}) and (\ref{eq:f23})] adjust
such that there is no net force (and hence no net movement of the membrane)
in the $z$-direction; we will here consider such incompressible fluids
and also assume the membrane to be incompressible. Nevertheless the
investigation of the net force $f^{(0)}=f_{1-2}^{(0)}+f_{2-3}^{(0)}$
on a membrane provides some insights into the electrostatic problem
under consideration here, and is given in appendix
\ref{sec:force_flat_membrane}.

Let us now proceed by considering the $z$-component of first order
correction to the forces,
$\bar{f}^{(1)}=\bar{f}_{1-2}^{(1)}+\bar{f}_{2-3}^{(1)}$; from
$\bar{f}^{(1)}$ one can obtain the work on the membrane under an 
undulation deformation of the shape, and thereby the free energy and
electrostatic contribution to membrane mechanical parameters (through
a power series in $q$, i.e. a long wavelength expansion). In
particular we want to compare the results of such an expansion to the
corresponding result for a ``free'' membrane: the free energy $G$ for
a membrane is described by the Helfrich form
\cite{helfrich73,seifert97} $G=\int d A \left( 2K H^2
+\sigma\right)\;$ where $H$ is the mean curvature, $d A$ the area
element on the membrane, $\sigma$ is the tension and $K$ is the
bending rigidity. The restoring force is then obtained as $f_{\rm
rs}=-\delta G/\delta h$ giving in $q$-space 
\begin{equation}
\bar{f}_{\rm rs}(q)=-\left[\sigma q^2+K q^4\right]{\bar h} +O({\bar h}^2).
\end{equation}
This type of expansion requires only that the expectation value
of $(\nabla h(x,y))^2$ is small \cite{seifert97}, i.e. that the
characteristic fluctuation amplitude is small compared to $1/q$.  In
the presence of an applied potential there will be electrostatic
contributions $\sigma_{\rm el}$ and $K_{\rm el}$ to the tension and
bending rigidity, so that $\sigma \rightarrow \sigma+\sigma_{el}$ and
$K \rightarrow K+K_{\rm el}$. Below we proceed by expanding
$\bar{f}^{(1)}$, using the results in Eqs. (\ref{eq:f12}) and
(\ref{eq:f23}), in a power series in $q$ for different limits in order
to obtain $\sigma_{\rm el}$ and $K_{\rm el}$ (note, however, in the
expansion for the dielectric limit, for $qL\gg 1$, we also find terms
odd in $q$). Note that since the tension and bending
rigidities are identified through terms in the restoring force
expansion which are proportional to $\bar{h}$, second and higher order
terms in $\bar{h}$ (see discussion at the end of the previous section)
do not contribute to $\sigma_{\rm el}$ and $K_{\rm el}$.

\begin{enumerate} 

\item
In the {\em ``small'' screening length} limit , $\kappa_1 L, \kappa_3 L
\gg 1$, a straightforward but lengthy expansion of $\bar{f}^{(1)}$ in a
power series in $q$ assuming that the wavelength of the perturbation
($=2\pi/q$) is larger than the membrane thickness and the Debye
screening lengths, $qd,q/\kappa_1,q/\kappa_3\ll 1$, gives
  \be
\bar{f}^{(1)}=-\left[\sigma_{\rm el} q^2+K_{\rm el} q^4 + O(q^6)\right]{\bar h}.\label{eq:f1_el}
  \ee
The explicit expression for the 
electrostatic contribution to the tension is
  \be
\sigma_{\rm el} = -  \frac{\varepsilon_0 \Delta \phi_{\rm m}^2 } {2}  \frac{  l_1 + l_3 + 4d/\varepsilon_2 }{(l_1 + l_3 + 2d/\varepsilon_2 )^2},\label{eq:Sigma}
   \ee
where, as before, the ``rescaled" Debye screening lengths are
   $l_1=(\varepsilon_1\kappa_1)^{-1}$ and
   $l_3=(\varepsilon_3\kappa_3)^{-1}$. We also introduced
  \bea
\Delta \phi_{\rm m}&=&\phi^0_1-\phi^0_3\nonumber\\
&=&\frac{\Delta\phi}{2}\cdot\frac{l_1+l_3+2 d /\varepsilon_2}{l_1+l_3+d/ \varepsilon_2}\label{eq:phi_m}
  \eea
being the potential difference between the main parts of the two bulk
fluids. Notice that $\sigma_{\rm el}$ gives a negative contribution to
the tension. The fact that $\sigma_{\rm el}<0$ originates from the
fact that that the applied potential creates a net charge density on
either side of the membrane surfaces, see Fig. \ref{fig:pot}; since
ions of equal charge repel each other the system would, for a
compressible membrane, be able to decrease the free energy by
separating the charges through a stretching of the membrane (i.e., by
increasing the membrane area). For an incompressible membrane (as
assumed here) the membrane is likely to respond to the
electrostatically induced negative tension by an opposite increase in
the membrane elastic contribution to the tension. If the magnitude
of $\sigma_{\rm el}$ exceeds the membrane elastic
strength (tensile strength) a {\em stretching instability}
occur. For the symmetric case ($\kappa=\kappa_1=\kappa_3$ and
$\varepsilon=\varepsilon_1=\varepsilon_3$) Eq. (\ref{eq:Sigma})
becomes:
\bea
\sigma_{\rm el}&=& \sigma_{\rm el}^0 \frac{1+  \tilde{\varepsilon}_m (\kappa d)^{-1}/2}{(1+\tilde{\varepsilon}_m (\kappa d)^{-1})^2}\nonumber\\
\sigma_{\rm el}^0&=& - \frac{\varepsilon_0 \varepsilon_2 (\Delta
  \phi_m)^2}{2d} \label{eq:Sigma_el_symm}
\eea
where we introduced the ratio $\tilde{\varepsilon}_m \equiv
\varepsilon_2/\varepsilon$ between the membrane and surrounding medium
dielectric constant (typically $\tilde{\varepsilon}_m\approx 1/40$, see
\cite{andelman95,lacoste06}). We notice that when $s=\tilde{\varepsilon}_m
(\kappa d)^{-1}\ll 1$, i.e. for an effectively small screening length
compared to the membrane width, the tension approaches $\sigma_{\rm
  el}^0$; the dimensionless parameter $s$ is commonly appearing in
membrane electromechanical problems, see
Refs. \cite{andelman95,lacoste06}. From Eq. (\ref{eq:Sigma}) we notice
that for the asymmetric case we similarly have $\sigma_{\rm el}\approx
\sigma_{\rm el}^0$ for $s_\gamma=(\varepsilon_2/\varepsilon_\gamma)
(\kappa d)^{-1}\ll 1$, where $\gamma=1,3$. Since $\sigma_{\rm el}^0$
only depend on the membrane dielectric constant, membrane width, and
the applied potential, $\sigma_{\rm el}$ is for large salt
concentration independent on the properties of the surrounding medium
(i.e. independent on $\kappa_1$, $\kappa_3$, $\varepsilon_1$ and
$\varepsilon_3$). The origin of this result is discussed below. 

The electrostatic contribution to the bending rigidity is
  \be
K_{\rm el}= \varepsilon_0 \Delta \phi_m^2 \frac{b_0 + b_1 d + b_2 d^2 + b_3 d^3 + b_4 d^4}{(l_1  + l_3+ 2d/\varepsilon_2)^3} \label{eq:K}
  \ee 
with coefficients
  \bea
b_0&=& \frac{1}{8} \Big( ( l_1 \kappa_1^{-2} + l_3 \kappa_3^{-2} )(l_1+l_3)\nonumber\\
& &\hspace{0.5cm} +2 l_1 l_3 (\kappa_1^{-1} +\kappa_3^{-1})^2 \Big), \nonumber\\
b_1&=& \frac{1}{4} \Big( \frac{3}{\varepsilon_2} (l_1 \kappa_1^{-2} + l_3 \kappa_3^{-2} )\nonumber\\
& &\hspace{0.5cm} + 8 l_1 l_3 (\kappa_1^{-1}+\kappa_3^{-1})  \Big),\nonumber\\
b_2&=& 2 \Big( \frac{1}{\varepsilon_2} (l_1 \kappa_1^{-1} + l_3 \kappa_3^{-1} ) +2 l_1 l_3 \Big),\nonumber\\
b_3&=& \frac{8}{3} \frac{1}{\varepsilon_2} (l_1+l_3),\nonumber\\
b_4&=& \frac{4}{3} \frac{1}{\varepsilon_2^2}.\label{eq:coeff_d_exp}
  \eea
We point out that $K_{\rm el}> 0$, i.e., the applied potential tends
to make the membrane more rigid towards bending. During a bending
deformation the induced charge density on one side of the membrane
gets compressed, whereas the charge density on the opposite side gets
expanded. The free energy changes of compression and expansion has
different signs, but are in general of different magnitude. It is only
for the case that all Debye screening charges are collapsed onto the
surfaces ($\kappa \rightarrow\infty$) and zero membrane thickness
$d\rightarrow 0$ that the expansion and compression free energies are
identical and $K_{\rm el}=0$ [see Eq. (\ref{eq:K})]. Thus, loosely speaking,
the smaller the ``effective'' membrane thickness (the membrane
thickness including the Debye screening layer thicknesses) the smaller
is the bending rigidity. For the symmetric case ($\kappa=\kappa_1=\kappa_3$ and
$\varepsilon=\varepsilon_1=\varepsilon_3$) we write Eq. (\ref{eq:K})
according to
\bea
K_{\rm el}&=&K^0_{\rm el} (1+\tilde{\varepsilon}_m (\kappa d)^{-1} )^{-3}\nonumber\\
& &\hspace{-1cm}\times \Big( 1+4 \tilde{\varepsilon}_m (\kappa d)^{-1} +
3\tilde{\varepsilon}_m (1+\tilde{\varepsilon}_m) (\kappa d)^{-2}\nonumber\\
& &\hspace{-1cm} + \tilde{\varepsilon}_m (9/8 + 3\tilde{\varepsilon}_m) (\kappa d)^{-3} +(9/8) \tilde{\varepsilon}_m^2 (\kappa d)^{-4} \Big) \nonumber\\
K^0_{\rm el}&=& \frac{1}{6} \varepsilon_0 \varepsilon_2 d (\Delta \phi_m)^2\label{eq:K_el_symm}
\eea
We note that in the limit of small relative membrane dielectric constant
$\tilde{\varepsilon}_m \rightarrow 0$ as well as for small
screening length compared to the membrane thickness, $(\kappa
d)^{-1}\rightarrow 0$, we have $K_{\rm el}\rightarrow K^0_{\rm
  el}$. Notice that, in practice, $\tilde{\varepsilon}_m$ is always
small $\approx1/40$, and therefore we have $K_{\rm el}\rightarrow
K^0_{\rm el}$ also for ``not too small'' values for $(\kappa d)^{-1}$.
Similarly, we note from Eq. (\ref{eq:K}) that for the asymmetric case
we have $K_{\rm el}\rightarrow K^0_{\rm el}$ in the
$\varepsilon_2/\varepsilon_\gamma \rightarrow 0$ and in the
$(\kappa_\gamma d)^{-1}\rightarrow 0$ ($\gamma=1,3$) limits. For the
above considered limits the major part of the potential drop is across
the membrane (since $\varepsilon_2/\varepsilon_\gamma$ or $(\kappa
d)^{-1}$ is small), and hence the electric field is essentially zero
everywhere except for the membrane region. Therefore, the membrane
parameters play the dominant role in the expression for $K_{\rm el}$
and $\sigma_{\rm el}$ (notice that $K^0_{\rm el}$ and $\sigma^0_{\rm
  el}$ depend only on $\varepsilon_0\varepsilon_2$, $d$ and $\Delta
\phi_m$), provided that the interactions of the induced surface
charges do not occur through the surrounding medium (again, certified
if $\varepsilon_2/\varepsilon_\gamma$ or $(\kappa d)^{-1}$ is
small). Given that $K_{\rm el}$ and $\sigma_{\rm el}$ can only depend
on $\varepsilon_0\varepsilon_2$, $d$ and $\Delta \phi_m$ in the limits
considered above one may use dimensional arguments to argue that
the limiting results, $K^0_{\rm el}$ and $\sigma^0_{\rm el}$, for the
bending rigidity and tension must (up to a constant prefactor) take the
forms given in Eqs. (\ref{eq:Sigma_el_symm}) and (\ref{eq:K_el_symm}).

Fig. \ref{fig:Klarge} illustrates the electrostatic contribution to
the tension and membrane bending rigidity and its dependence on salt
concentration (Debye screening) for the symmetric case for
simplicity. We see that the absolute value of electrostatic
contribution to the membrane tension increases with increasing salt
concentration, i.e. for increasing $\kappa$. We attribute this to an
increase of screening charges (in a layer of decreased thickness) next
to the membrane; the increased amount of charges will result in larger
electrostatic repulsion between ions in the screening clouds [see
discussion following Eq. (\ref{eq:phi_m})]. In contrast to the effect on
tension the electrostatic contribution to the bending rigidity
decreases with increasing salt concentration (with $K_{\rm el}$
approaching $K^0_{\rm el}$ as $\kappa\rightarrow \infty$). The reason
behind this is that for bending properties the thickness of the Debye
screening layers plays a role - a larger ``effective'' membrane
thickness gives a higher bending rigidity [see discussion following
Eq. (\ref{eq:K}), and the spontaneous curvature calculation in
Appendix \ref{sec:alternative}]. If we choose the potential difference
between the membrane sides $\Delta \phi_m =$ 100 mV, $\varepsilon_2=2$
and $d=2.5$ nm we find that $K^0_{\rm el}=0.018 k_B T$ (for room
temperature, $k_B T=4\cdot 10^{-21}$). From Fig. \ref{fig:Klarge} we
see that $K_{\rm el}/K^0_{\rm el}$ can become quite large for small
$\kappa$ and, therefore, for sufficiently small salt concentration the
bending rigidity $K_{\rm el}$ can exceed the thermal energy $k_B T$;
we therefore expect that the increase of the bending rigidity in the
presence of an applied electrostatic potential predicted in this study
can indeed be experimentally observed for sufficiently large $\Delta
\phi_m$ and small salt concentrations (note however, the below
restriction on salt concentration due to assumptions in the
Debye-H\"uckel approximation).

Let us compare the results above for the symmetric case to the results
obtained in \cite{lacoste06}. We find that the result for $\sigma_{\rm
  el}$ given in Eq. (\ref{eq:Sigma_el_symm}) agrees with the finite
bilayer thickness, non-conductive membrane tension (there denoted by
$\Sigma_{\rm in}+\Sigma_{\rm out}$) obtained in \cite{lacoste06} (note
that in \cite{lacoste06} the membrane thickness is denoted by $d$,
whereas we denote by $d$ the size of a lipid monolayer, so that in our
case the membrane thickness is $2d$. Also note that, due to different
boundary conditions at the electrodes, the $V$ in \cite{lacoste06}
should be equated with our $\Delta\phi_m$). Concerning the bending
rigidity result, we note that no explicit expression for $K_{\rm el}$
was given in \cite{lacoste06}, only a numerical value for a specific
set of parameter values. We choose the same parameter values ($\Delta
\phi_m=50$ mV, $d=2.5$ nm, $\tilde{\varepsilon}_m=1/40$, and
$2d\kappa=7.4$), but note that in order to get the expression for
$K_{\rm el}$ one must also choose the actual value of $\varepsilon_2$
(not just the ratio $\varepsilon_2/\varepsilon$), which was not
specified in \cite{lacoste06}. We choose the standard value
$\varepsilon_2=2$ \cite{andelman95,sens02} and then find that $K_{\rm
  el}=0.00467\,k_B T$, which is a bit less than half the value found
in Ref. \cite{lacoste06}.  Since no explicit expression for $K_{\rm
  el}$ was given in \cite{lacoste06} it is difficult to comment on the
nature of this discrepancy. Finally, using the approximate expression
$K_{\rm el}\approx K^0_{\rm el}$ for the parameters above we find that
this approximation underestimates $K_{\rm el}$ by merely 1 $\%$.

Here, a few words on the validity of the Debye-H\"uckel approximation,
used throughout this study, are in place.  This approximation should
work when the quantity $I=\beta q^i
(\Phi_\gamma(\vec{x})-\phi^0_\gamma)$, where $\gamma=1,3$, is very
small, i.e. $I\ll 1$ (see Sec. \ref{sec:general_formulation}), but in
practice the Debye-H\"uckel approximation works well whenever $I<1$,
see Ref \cite{Hunter}. The maximum of $I$ occurs at the membrane
surfaces, see Eqs. (\ref{eq:Phi_flat_1}), (\ref{eq:Phi_flat_3}) and
Fig. \ref{fig:pot}, and we find that for the $\kappa_\gamma L\gg 1$
limit considered here we have $I=\beta q^i \Delta \phi l_\gamma
/[2(l_1+l_3+d/\varepsilon_2)]$. Using $d\approx 2.5$ nm,
$\varepsilon_2\approx 2$ and $\varepsilon_\gamma\approx 80$, the
denominator in the expression for I above is dominated by the
$d/\varepsilon_2$ term whenever $\kappa_\gamma^{-1} < 50$ nm (in this
limit also $\Delta \phi \approx \Delta \phi_m$); for such ion
concentrations we have the following criterion for the validity of the
Debye-H\"uckel approximation
  \be
I= \frac{1}{2} \beta q^i \Delta \phi_m  s_\gamma < 1.
  \ee
again involving the parameter
$s_\gamma=(\varepsilon_2/\varepsilon_\gamma)(\kappa_\gamma
d)^{-1}$. For a large potential difference $\Delta \phi_m =$ 100 mV we
find that $I<1$ when $\kappa_\gamma^{-1} < 50$ nm (using
$q^i=1.6\times 10^{-19}\,{\rm C}$ and assuming room temperature). This
means that the Debye-H{\"u}ckel approximation, which was made in
Sec. \ref{sec:general_formulation} in the main text, works
surprisingly well in general; the reason for this is the small value
of $\varepsilon_2/\varepsilon_\gamma$ ($\approx 1/40$) guaranteeing
that the major part of the potential drop occurs across the membrane
and that, therefore, the potential drop across the electrolytes, for
which we applied the Debye-H\"uckel approximation, is modest, see
Fig. \ref{fig:pot}.  We point out that the dielectric limit,
$c^i_{{\rm bulk},\gamma}\rightarrow 0$, does not rely on a
Debye-H\"uckel approximation and results to be given below are
therefore valid for any value of the applied potential. We have
shown above that our results for the small screening length limit (with the
above explicit restriction) and in the dielectric limit are usually
valid. We note, however, that there is in general an intermediate salt
regime where the Debye-H\"uckel approximation breaks down (for
sufficiently large applied potentials). We leave the investigation of
this intermediate regime for future studies.

\begin{figure}
\begin{center}
\includegraphics[width=9cm]{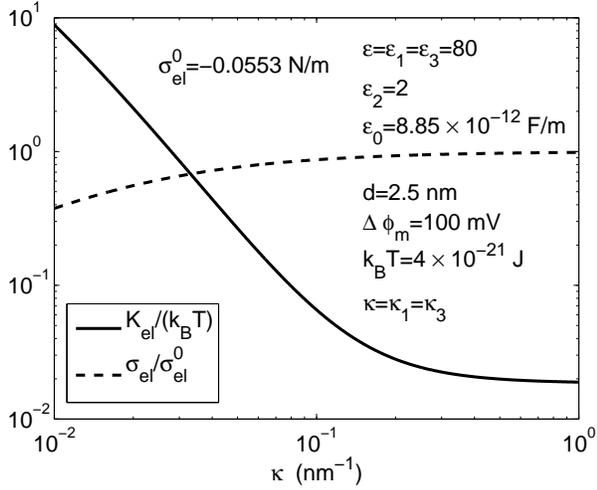}
\end{center}
\caption{The electrostatic contributions $\sigma_{\rm el}$ and $K_{\rm
el}$ to the tension and the bending rigidity and as a function of
Debye screening in the small screening length limit (symmetric case),
using Eq. (\ref{eq:Sigma_el_symm}) and (\ref{eq:K_el_symm}). We
express $K_{\rm el}$ in terms of the (room temperature) thermal energy
$k_BT$ and $\sigma_{\rm el}$ in terms of its infinite
$\kappa$ value $\sigma_{\rm el}^0$. The parameters used are
listed in the figure. Notice that increasing salt concentration,
i.e. increased $\kappa$, leads to a decrease in the electrostatic
contribution to the bending rigidity. In contrast, for increasing
$\kappa$ the magnitude of the electrostatic contribution to the
tension increases.}
\label{fig:Klarge}
\end{figure}

We finally point out that there are alternative ways of computing the membrane
mechanical parameters.  For instance one can calculate the tension by
finding the integral of the deviation of the pressure profile from the
value of the pressure far from the membrane. This approach is
demonstrated in Appendix \ref{sec:alternative}, giving the same result
as Eq. (\ref{eq:Sigma}) for the tension. In that appendix the same
type of approach is also used in order to obtain the electrostatic
contribution to the membrane spontaneous curvature, see
Eq. (\ref{eq:C0}).

\item in the {\em dielectric limit}, $\kappa_1,\kappa_3\rightarrow 0$,
  one can again perform a power series expansion in $q$ using
  Eqs. (\ref{eq:f12}) and (\ref{eq:f23}) and assuming $qd\ll 1$. In
  addition, there are two limits of interest depending on whether the
  wavelength perturbation is smaller or larger than $L$.

For (i) $q L\gg 1$ we find that 
  \be
\bar{f}^{(1)}=-[aq+\sigma_{\rm el}q^2+bq^3+K_{\rm el} q^4+O(q^5)]{\bar h},\label{eq:f1_el2}
  \ee
where 
  \be
a=-\varepsilon_0 \left( \frac{\Delta \phi}{L} \right)^2 \frac{(\varepsilon_1^{-1}-\varepsilon_3^{-1})^2}{(\varepsilon_1^{-1}+\varepsilon_3^{-1})^3}\label{eq:a}
  \ee
and
\bea
\sigma_{\rm el}&=&-8 d\varepsilon_0 \varepsilon_2 \left( \frac{\Delta \phi}{L} \right)^2 \frac{\varepsilon_1^{-1}\varepsilon_3^{-1}}{(\varepsilon_1^{-1}+\varepsilon_3^{-1})^4}\nonumber\\
& &\times (\varepsilon_1^{-1}-\varepsilon_2^{-1})(\varepsilon_3^{-1}-\varepsilon_2^{-1}),\label{eq:Sigma_diel}
\eea
and the higher order terms in $q$ are more complicated functions of
the dielectric constants and $d$. The term linear in $q$ is negative
and is expected to cause an instability for long
wavelengths \cite{note2}. We notice that the prefactor $a$ [see
Eq. (\ref{eq:a})] is proportional to the membrane polarization charge
density (for a flat membrane) squared, $a\propto (\rho^s)^2$, see
Eq. (\ref{eq:rho_s}). The linear, non-analytic, $q$ term may be
interpreted as follows: for the asymmetric dielectric case the
external potential induces an effective net membrane polarization
charge density $\rho^s$. The membrane charges interact via the
unscreened (there are no ions in the present limit) Coulomb
interaction, giving rise to the $(\rho^s)^2$ proportionality for $a$;
the non-analyticity of the free energy arises due to the long-range
character of the Coulomb interaction (which decays as $1/r$, where $r$
is the distance between charges). We point out that the linear $q$
term does not depend on the membrane parameters, $d$ and
$\varepsilon_2$, which means that this instability should exist for
any interface between coexisting fluids (fluid interface instabilities
of a rather similar nature has been investigated previously, see for
instance \cite{melcher65}). For the symmetric case
$\varepsilon_1=\varepsilon_3$ we see that $a=0$, but the third order
term above is still present, $b\neq 0$, in general (see discussion
below).

For the case (ii) $qL\ll 1$, i.e. the wavelength perturbation is
longer than the distance between the electrodes, the first order force
takes the form $\bar{f}^{(1)}=-[c+\sigma_{\rm el}q^2+K_{\rm el}
  q^4+O(q^6)]{\bar h}$. In this limit the interactions between the
 membrane polarization charges become short-range due to
effectively small (compared to the wavelength of the perturbation)
distance of the membrane to the electrodes. In this limit we thus do
not have any odd $q$ terms the effect of the applied potential is
simply to give an electrostatic contribution to the tension and
bending rigidity. The explicit expressions become somewhat
complicated, but can be produced straightforwardly by a small $q$
expansion using a symbolic mathematical software like Mathematica or
Maple together the expressions for $\bar{f}_{1-2}$ and $\bar{f}_{2-3}$
given in Eqs. (\ref{eq:f12}) and (\ref{eq:f23}).

\item For {\em the symmetric case}
($\varepsilon=\varepsilon_1=\varepsilon_3$,
$\kappa=\kappa_1=\kappa_3$) the lowest order term in the $q$ expansion
is the tension term ($q^2$ term), both in the small screening length
limit and for the dielectric limit. For both these limits we have that
the electrostatic contribution to the tension $\sigma_{\rm el}$ is
negative. Thus when $|\sigma_{\rm el}|$ becomes sufficiently large,
i.e.  large applied potential, a membrane stretching instability can
occur. We also find that in the symmetric dielectric limit there is a $q^3$
term present (for $qL\gg 1$), where explicitly we find [see
Eq. (\ref{eq:f1_el2})]
\be
b= 2 \varepsilon_0 \varepsilon \varepsilon_2^2 (\frac{\Delta \phi}{2L})^2  (\varepsilon^{-1}-\varepsilon_2^{-1})^2 d^2. 
  \ee
We notice that $b\propto P_z^2$, where $P_z$ is the membrane
polarization per unit area, see Eq. (\ref{eq:P_z}). For the symmetric
dielectric case the effect of the applied potential is to polarize the
membrane and the $q^3$-term is expected to originate from induced,
unscreened, dipole-dipole interactions (which decays as $1/r^3$, where
$r$ is the distance between the dipoles \cite{Jackson}) within the
membrane.

\end{enumerate}

\section{Summary and discussion}\label{sec:summary_outlook}

We have in this paper derived expressions for the electrostatic
contributions to biomembrane mechanical parameters (such as tension
and bending rigidity) in the presence of an static applied potential across a
membrane. The membrane was assumed non-compressible, non-conductive
(membrane region described by Laplace equation) and surrounded by
electrolyte solutions (described by the Debye-H\"uckel equation). By
solving the equations for the electrostatic potential and using the
stress-tensor the forces acting on the membrane were obtained, which
in turn were used to obtain the free energy and the electrostatic
contribution to the membrane mechanical parameters as a function of
the applied potential, the salt concentrations (entering through the
Debye screening lengths) and the dielectric constants of the membrane
and the solvents. Results of particular interest, that are found in
this study, include: for (1.) the {\em small screening length limit},
where the Debye screening length is smaller than the distance between
the electrodes, the screening certifies that all electrostatic
interactions are short-range, leading to a free energy expansion of
the form $\sim \sigma_{\rm el}q^2+K_{\rm el}q^4+O(q^6)$ (where $q$ is
the wavevectors), the main effects of the applied potential are to
decrease the membrane tension and increase the bending rigidity;
explicit expression are given in Eqs. (\ref{eq:Sigma}) and
(\ref{eq:K}). Our expression for the tension for the symmetric case
reproduces the result in \cite{lacoste06}. In \cite{sens02} it was
also found that an applied electric field gives a negative
contribution to the tension. However in that study the medium
surrounding the membrane was characterized by conductivities rather
than Debye screening lengths, and it is therefore difficult to
directly compare our results to theirs. For sufficiently large applied
potentials the magnitude of the electrostatic contribution to the
tension will exceed the maximum tension the membrane can sustain,
leading to a membrane stretching instability. Possibly, this
instability can result in the formation of pores and flow of ions
through the membrane (in fact, the membrane tension is one of the key
parameters in the modeling of membrane electroporation dynamics
\cite{joshi02}). For (2.) the dielectric limit, i.e. no salt (for
small wavevectors $q$ compared to the distance between the
electrodes), when the dielectric constants on the two sides are
different, the applied potential induces an effective (unscreened)
membrane charge density, whose long-range interaction causes a
membrane undulation instability if the dielectric constants of the two
bulk fluids are different; this effect is characterized by a negative term
linear in $q$ in the free energy expansion, see Eqs
(\ref{eq:f1_el2}) and (\ref{eq:a}), i.e. this term is of lower order
in $q$ than the tension term. Previous similar results include: In
\cite{melcher65} the interface between two immiscible fluids of
different dielectric constants was found to be unstable in the
presence of a perpendicular electric field. The case of stiff
(charged) DNA with bound, but mobile, counter ions was investigated in
\cite{Golestanian} and a shape instability found (supposedly leading
to a DNA condensation). If the two dielectrics on each side of the
membrane are identical, we found that then the applied electric field
will give a negative contribution to the tension. Hence, if the
applied potential is sufficiently large a membrane instability occurs
also for the (dielectric) symmetric case.

We quantified the validity of the Debye-H\"uckel approximation (used
throughout this study) and showed that our results are in general
valid in the small screening length limit as well as in the
dielectric limit. However for ``small'', but non-zero, salt
concentration and large applied potential the Debye-H\"uckel
approximation is no longer valid and one needs to consider the full
Poisson-Boltzmann equation. It remains a future challenge to solve the
full Poisson-Boltzmann problem in order to find expressions for the
free energy for arbitrary salt concentration, and, in particular, to
investigate in more detail the nature of the onset of the predicted
membrane instability (via the negative term linear in $q$ in the free
energy) as salt concentration is lowered.

Possible applications of the result for the small screening length
limit above would be to lipid membranes where a potential difference
across the membrane is enforced by ion pumps incorporated in the
membrane. Changes in membrane rigidity might then be observed in
micropipette or video microscopy experiments, if the screening length
and the membrane potential are large, see Fig.
\ref{fig:Klarge}. Another possible experiment would be to observe the
structural change of domains in a multi-component membrane using
fluorescence correlation spectroscopy (FCS) as the membrane potential
in a patch clamp experiment is altered; one might be able to observe a
change from a phase with caps to one with stripes or buds as the
effective bending rigidity is changed by the applied potential
\cite{harden05}.

For the cases discussed above where a membrane instability occurs, the
system is expected to be driven from the quasi-flat shape into a new
equilibrium configuration. Our perturbation analysis cannot in general
say anything about this new configuration. However, for the small
screening length limit we above speculated that the negative
electrostatic contribution to the tension (for large applied
potentials) could lead to electroporation and a corresponding flux of
ions through the membrane. It may also be speculated (similar to the
studies in Refs. \cite{Kim_Sung} and \cite{sens02}) that the
electrostatically induced new equilibrium configuration under certain
conditions could be a spherical membrane (a vesicle); it is known that
vesicles can be created in laboratories in a process known as
electroformation \cite{angelova86} under the application of electric
fields.  We hope that the present theory will stimulate further work
directed towards the controlled estimation of vesicle sizes as a
function of electrostatic parameters. e.g. potential differences and
electrolyte concentration.

\begin{acknowledgments}
We would like to thank John H. Ipsen and James L. Harden for valuable
discussions. We also acknowledge our referees for useful comments.
\end{acknowledgments}

\appendix

\section{Derivation of bulk equations}\label{sec:derivations}

In this appendix we show how the electrostatic equations and the
stress-tensor used in the main text can be derived as Euler-Lagrange
equations from a free energy. Suppressing the region index $\gamma$
for convenience we can write the appropriate free energy as
\begin{eqnarray}
G&=&\int d^3x\;\bigg[-\frac{1}{2}\varepsilon_0\varepsilon|\vec{\nabla}\Phi(\vec{x})|^2+\rho\Phi-p\nonumber\\
&&+\sum_i c^i\left(k_B T\left(\ln\frac{c^i}{c_{{\rm Total}}}-1\right)+\mu^i\right)\bigg].
\end{eqnarray}
Here $p=p(\vec{x})$ is the local pressure, $c^i=c^i(\vec{x})$ are the
local concentrations of the ions, $c_{\rm Total}=c_{\rm
  Total}(\vec{x})$ is the total concentration of molecules of all
kinds (including water) and $\mu^i$ are (constant) chemical potentials
for the ions. The logarithmic term corresponds to the entropy of
mixing. Some of the Euler-Lagrange equations of this free energy can
be found by demanding stationarity when varying the ion concentrations
\begin{equation}
0=\left.\frac{\delta G}{\delta c^i}\right|_{c_{{\rm Total}}}=\Phi q^i+k_B T\ln\frac{c^i}{c_{{\rm Total}}}+\mu^i.
\end{equation}
Solving for the ion concentrations we find
\begin{equation}
c^i=c_{{\rm Total}}\exp\left(-\frac{q^i\Phi+\mu^i}{k_B T}\right).\label{eq:ionconc}
\end{equation}
Inspection of the expression for the charge density $\rho=\sum_i q^i
c^i$ with $c^i$ given by Eq. (\ref{eq:ionconc}) reveals that there
is a unique value of $\Phi$ for which $\rho$ vanishes. We define
$\phi^0$ to be this ``equilibrium value", i.e.
\begin{equation}
\left.\rho\right|_{\Phi=\phi^0}=0.
\end{equation}
The ion concentrations at ``equilibrium" is labeled by
\begin{equation}
c^i_{{\rm bulk}}=\left.c^i\right|_{\Phi=\phi^0}.\label{eq:cbulk}
\end{equation}
The Euler-Lagrange equation for $\Phi$ is
\begin{equation}
0=\frac{\delta G}{\delta \Phi}=\varepsilon_0\varepsilon\vec{\nabla}^2\Phi+\rho,\label{eq:gausslaw}
\end{equation}
which is simply the Poisson equation. Insertion of
Eq. (\ref{eq:ionconc}) and Eq. (\ref{eq:cbulk}) gives the
Poisson-Boltzmann equation
\begin{equation}
\varepsilon_0\varepsilon\vec{\nabla}^2\Phi+\sum_i q^i c^i_{{\rm bulk}}\exp\left(-\frac{q^i(\Phi-\phi^0)}{k_B T}\right)=0,\label{eq:PB}
\end{equation}
which when linearized gives the Debye-H{\"u}ckel equation,
Eq. (\ref{eq:DH}) in the main text.

The force-balance in the bulk regions can be found by demanding that
$G$ should be stationary with respect to moving all fluid elements,
particles and fields by an infinitesimal position dependent distance
$\delta\vec{x}=\delta\vec{x}(\vec{x})$. Denoting fields after the move by a
prime one has the new fields
\begin{eqnarray}
\Phi'(\vec{x}')&=&\Phi(x),\\
p'(\vec{x}')&=&p(x),\\
{c^i}'(\vec{x}')&=&(1-\vec{\nabla}\cdot\delta\vec{x})c^i(\vec{x}),
\end{eqnarray}
where $\vec{x}'=\vec{x}+\delta\vec{x}$. The free energy after the move is
\begin{eqnarray}
G'&=&\int d^3x'\;\bigg[-\frac{1}{2}\varepsilon_0\varepsilon|\vec{\nabla}'\Phi'(\vec{x}')|^2+\rho'\Phi'-p'\nonumber\\
&+&\sum_i {c^i}'\left(k_B T\left(\ln\frac{{c^i}'}{c_{{\rm Total}}'}-1\right)+\mu^i\right)\bigg],
\end{eqnarray}
and using
\begin{eqnarray}
d^3x'&=&(1+\vec{\nabla}\cdot\delta\vec{x})d^3x,\\
\frac{\partial}{\partial x_i'}&=&\sum_j\left(\delta_{ij}-\frac{\partial \delta
    x_j}{\partial x_i}\right)\frac{\partial}{\partial x_j},
\end{eqnarray}
one finds that the change in free energy is
\begin{equation}
\delta G=G'-G=\int d^3x\;\sum_{i,j}T_{ij}\frac{\partial\delta x_j}{\partial
  x_i},\label{eq:deltaG}
\end{equation}
where $T_{ij}$ is the stress tensor for the system
\begin{equation}
T_{ij}=\varepsilon_0 \varepsilon \left( \frac{\partial \Phi}{\partial x_i}\frac{\partial \Phi}{\partial x_j}-\frac{1}{2} \delta_{ij} \sum_k  \frac{\partial \Phi}{\partial x_k}\frac{\partial \Phi}{\partial x_k}\right)-p\delta_{ij}.\label{eq:stresstensor}
\end{equation}
Note that the first two terms on the right hand side of
Eq. (\ref{eq:stresstensor}) are simply the classic Maxwell stress
tensor. Performing a partial integration in Eq. (\ref{eq:deltaG}) one can see
that the condition of stationarity of G implies the conservation of stress
\begin{equation}
\sum_i \frac{\partial T_{ij}}{\partial x_i}=0.
\end{equation}
Combing this with Eq. (\ref{eq:gausslaw}) one arrives at the more physically revealing form
\begin{equation}
-\rho\vec{\nabla}\Phi-\vec{\nabla}p=0,\label{eq:forcebalance}
\end{equation}
i.e. electric forces should be balanced by changes in hydrostatic
pressure. Finally, since we know the behavior of the charge density $\rho$ as a
function of $\Phi$ from Eq. (\ref{eq:ionconc}) we can integrate
Eq. (\ref{eq:forcebalance}) to find the pressure as a function of
$\Phi$
\begin{equation}
p=p^0+k_B T\sum_i c^i_{{\rm bulk}}\left[\exp\left(-\frac{q^i(\Phi-\phi^0)}{k_B T}\right)-1\right].\label{eq:pressure}
\end{equation}
The quadratic version of Eq. (\ref{eq:stresstensor}) with
Eq. (\ref{eq:pressure}) inserted is Eq. (\ref{eq:sigmaij}) of the main
text. Note that the pressure can also be written $p=p^0+k_B T\sum_i
(c^i-c^i_{{\rm bulk}})$, i.e. there is an osmotic contribution to the
pressure when ions are ``confined" by the electric potential (see for instance
\cite{israelachvili}).

\section{Total net force on a flat membrane}\label{sec:force_flat_membrane}

In this appendix we investigate the total force acting on a {\em
 flat} membrane in some detail.

From Eqs. (\ref{eq:f12}), (\ref{eq:f23}) and
Eq. (\ref{eq:coeff_flat_1}) we obtain the following explicit
expression for the total force (per unit area) acting on a {\em flat}
membrane:
  \bea
f^{(0)}&=&f_{1-2}^{(0)}+f_{2-3}^{(0)}\nonumber\\
&=&p^0_1-p^0_3\nonumber\\
& &\hspace{-1.5cm}-\frac{\varepsilon_0}{2} (\Delta \phi)^2 \Big(\frac{\varepsilon_1^{-1} e^{-\kappa_1L}}{(1+e^{-\kappa_1L})^{2}}-\frac{\varepsilon_3^{-1} e^{-\kappa_3L}}{ (1+e^{-\kappa_3L})^{2}}\Big) \Gamma^2
\label{eq:total_force}
 \eea
and $\Gamma= 1/[ g(\kappa_1) l_1 + g(\kappa_3)l_3 +d/\varepsilon_2] $,
and $l_1=(\varepsilon_1\kappa_1)^{-1}$ and
$l_3=(\varepsilon_3\kappa_3)^{-1}$ as before [the function $g(q)$ is
defined in Eq. (\ref{eq:g_q})].  If the fluids surrounding the
membrane are incompressible, then the pressures $p^0_\gamma$ adjust
such that the above force vanishes. In fact, in this study we assume that both the
membrane and the surrounding fluids are incompressible. Let us,
however, in order to gain some physical insights, investigate
different limits of the non-pressure part of the above force (i.e. we
take $p^0_1=p^0_3$). 

\begin{enumerate}

\item In {\em the ``small'' screening length limit}, $\kappa_1 L,
\kappa_3L \gg 1$, the non-pressure part of Eq. (\ref{eq:total_force})
becomes:
  \be
f^{(0)}\approx 0. \label{eq:total_force1}
  \ee
The force is zero whenever the screening length is small compared to
the distance between the electrodes. Thus it is not possible to get a
net force on the membrane solely by having different concentration of
ions (free charges) on the two sides. That there is no net force on
the membrane can be related to the fact that their is no net free
charge around the membrane, which can be understood from Gauss' law,
$\int \vec{D}\cdot \hat{n} dS=Q_{\rm free}$, where $\vec{D}$ is the
displacement field and $Q_{\rm free}$ is the enclosed free charge:
since far from the membrane on both sides the electric field is zero
(in the here considered limit, see Eqs. (\ref{eq:Phi_flat_1}),
(\ref{eq:Phi_flat_3}) and Fig. \ref{fig:pot}) so is the displacement
field, and applying Gauss law (for a large ``pillbox'' enclosing the
membrane and the Debye screening layers) one finds that $Q_{\rm
free}=0$, as well as no net charge due to changes in polarization of the
dielectric media. The effective net charge for the membrane and the
Debye screening layers is hence zero and there will be no net force
due to these charges.

\item In {\em the dielectric limit}, $\kappa_1,\kappa_3\rightarrow 0$,
  we find
  \bea
f^{(0)}&=&-2\varepsilon_0 (\frac{\Delta \phi}{2L})^2\nonumber\\
& &\hspace{-0.3cm}\times \frac{\varepsilon_1^{-1}-\varepsilon_3^{-1}}{\Big(\varepsilon_1^{-1}+\varepsilon_3^{-1}+2( d/L) \varepsilon_2^{-1}\Big)^2}\label{eq:total_force2}
  \eea
for the non-pressure part of the force. When there are no free
charges, as considered here, the Gauss law argument above does not
apply. One then needs to also take into account the bound charges,
which create a polarization surface charge density
$\rho^s_{1-2}=-(\vec{P}_2-\vec{P}_1)\cdot \hat{z}$ (see
\cite{Jackson}, chap. 4) on the region 1-2 interface. Similarly, for
the region 2-3 interface we have a polarization surface charge density
$\rho^s_{2-3}=(\vec{P}_2-\vec{P}_3)\cdot \hat{z}$. From the fact that
$\vec{P}=\vec{D}-\varepsilon_0 \vec{E}$ and that the normal component
of the $\vec{D}$-field is continuous we find that
$\rho^s_{1-2}=\varepsilon_0 (\vec{E}_2-\vec{E}_1)\cdot \hat{z}$ and
$\rho^s_{2-3}=-\varepsilon_0 (\vec{E}_2-\vec{E}_3)\cdot \hat{z}$,
i.e. the polarization surface charge density is determined by the jump
in the electric field at the interface. Using the explicit expressions
for the potential $\Phi(z)$ given in Sec. \ref{sec:flat_membrane} and that z-component of the electric field is $E_z=-\partial \Phi
/\partial z$ we find 
$\rho^s_{1-2}\approx -\varepsilon_0 E_{z,{\rm appl}}
(\varepsilon_1^{-1}-\varepsilon_2^{-1})/(\varepsilon_1^{-1}+\varepsilon_3^{-1})$
and $\rho^s_{2-3}\approx \varepsilon_0 E_{z,{\rm
appl}}(\varepsilon_3^{-1}-\varepsilon_2^{-1})/(\varepsilon_1^{-1}+\varepsilon_3^{-1})$,
with the applied field being $E_{z,{\rm appl}}=\Delta \phi/(2L)$ where we
assumed a small membrane thickness $d/L\ll 1$. The effective membrane
charge density $\rho^s=\rho^s_{1-2}+\rho^s_{2-3}$ then becomes
proportional to the dielectric asymmetry between region 1 and region
3, explicitly 
  \be
\rho^s\approx \varepsilon_0 E_{z,{\rm
appl}}\frac{\varepsilon_3^{-1}-\varepsilon_1^{-1}}{\varepsilon_1^{-1}+\varepsilon_3^{-1}}\label{eq:rho_s}
  \ee
where the proportionality $\rho^s\propto
\varepsilon_3^{-1}-\varepsilon_1^{-1}$ follows directly from the fact
that the jump in the electric field is proportional to
$\varepsilon_2/\varepsilon_1$ ($\varepsilon_2/\varepsilon_3$) for the
region 1-2 (region 2-3) interface (see discussion at the end of
Sec. \ref{sec:flat_membrane} and Fig. \ref{fig:pot}). Thus, whenever
$\varepsilon_1\neq \varepsilon_3$ we have that the membrane has an
effective net charge of induced bound charges, which when put into the
applied electric field $E_{z,{\rm appl}}$ gives rise to the force
above; note that the force is proportional to $\rho^s$ as it should
(for $d/L\ll 1$). We also note that Eq. (\ref{eq:total_force2}) for $d\to 0$ can be written
\begin{equation}
f^{(0)}=\frac{1}{2}\left(\frac{\varepsilon_1}{\varepsilon_3}-1\right)\varepsilon_0\varepsilon_1 \left(\hat{z}\cdot\vec{E}_1\right)^2\;,
\end{equation}
which can be compared with for instance the formula in \cite{landau} for the
pressure decrease in a fluid at the fluid-air interface due to a normal
electric field
\begin{equation}
\hat{z}\cdot\vec{E}_1=\frac{\Delta\phi}{2 L}\frac{2\varepsilon_1^{-1}}{\varepsilon_1^{-1}-\varepsilon_3^{-1}}
\end{equation}
on the fluid side.

\item For {\em the symmetric case} ($\varepsilon_1=\varepsilon_3=\varepsilon$,
$\kappa_1=\kappa_3$ and $p^0_1=p^0_3$) the non-pressure part of
Eq. (\ref{eq:total_force}) is zero,
  \be
f^{(0)}=0,
  \ee
since for the symmetric case there is no net surface charge, and hence no
net force on the flat membrane. We point out, however, that the
membrane gets polarized. In particular, the
polarization per unit area in the $z$-direction (using the results
above) is
  \bea
P_z&=&(-d)\rho^s_{1-2}+d\rho^s_{2-3}\nonumber\\
&=&\varepsilon_0\varepsilon E_{z,{\rm appl}}
(\varepsilon^{-1}-\varepsilon_2^{-1})  d\label{eq:P_z}
  \eea
in the dielectric limit.

\end{enumerate}

\section{Alternative approach to tension and spontaneous curvature}\label{sec:alternative}

In this appendix we will demonstrate an alternative approach to
deriving the tension in the membrane, which can also be used to obtain
the spontaneous curvature induced by the electric field. This approach
consists of calculating moments of the deviation of the pressure
profile from the value of the pressure far from the membrane. We will
therefore in this appendix only study the small screening length limit
where $L\to\infty$, such that the pressure approaches a well defined
pressure away from the membrane.

The tension is obtained as the integral of the lateral pressure
profile deviation, or equivalently the excess lateral stress, of the
planar membrane \cite{lomholt06c}. Choosing the diagonal $x$-component
of the stress tensor to represent the lateral stress one has the
precise formula \cite{lomholt06c}
\begin{equation}
\sigma=\int_{-\infty}^\infty d z\;\left[T^{(0)}_{x x}-\left(-p^0\right)\right]\;,
\end{equation}
where we have used that the
pressure on the two sides of the membrane should be identical
$p^0=p^0_1=p^0_3$, since the zeroth order force in the small screening
length limit, Eq (\ref{eq:total_force1}), vanishes. In terms of the
zeroth order solution given in Eqs. (\ref{eq:Phi_flat_2}),
(\ref{eq:Phi_flat_1}) and (\ref{eq:Phi_flat_3}) we obtain
\begin{eqnarray}
\sigma&=&-\frac{\varepsilon_0}{2}\varepsilon_1\kappa_1 A_1^2-\frac{\varepsilon_0}{2}\varepsilon_3\kappa_3 A_3^2-\varepsilon_0\varepsilon_2 d A_2^2\nonumber\\
&&-2 d \left(p^0_2-p^0\right)\;.\label{eq:tension_v2}
\end{eqnarray}
To obtain the final result we need to find $p^0_2$. To do this we note
that in a flat equilibrium configuration the force on the two region
boundaries should vanish, $f^{(0)}_{1-2}=0$ and $f^{(0)}_{2-3}=0$ (or
equivalently: $T_{zz}^{(0)}(z)=-p^0$ for all $z$). From either
Eq. (\ref{eq:f12}) or Eq. (\ref{eq:f23}) one finds
\begin{equation}
p^0_2=p^0+\frac{1}{2}\varepsilon_0\varepsilon_2A_2^2\;.
\end{equation}
This gives
\begin{equation}
\sigma=-\frac{\varepsilon_0}{2}\varepsilon_1\kappa_1 A_1^2-\frac{\varepsilon_0}{2}\varepsilon_3\kappa_3 A_3^2-2\varepsilon_0\varepsilon_2 d A_2^2\;.\label{eq:tension_v3}
\end{equation}
This expression is identical to the tension given by Eq. (\ref{eq:Sigma}) in the main text.

An advantage of the approach of integrating the stress profile is that
we can obtain the change in spontaneous curvature $C_0$ induced by the
electric potential. If we include a spontaneous curvature in the
Helfrich free energy $G$ given at the beginning of
Sec. \ref{sec:free_energy}, writing it as
\begin{equation}
G=\int d A \left[\frac{1}{2}K(2 H)^2 -K C_0 2 H+\sigma\right]\,
\end{equation}
and, as before, calculate the force $f_{\rm rs}=-\delta G/\delta h$,
then the spontaneous curvature will drop out at linear order in $h$
and we would end up with our previous expression for the force where
$C_0$ is not present; thus it is not possible to obtain the spontaneous
curvature using the approach of the main text. However, just like for
the tension, the spontaneous curvature can be obtained from the
lateral stress profile, namely as the negative first moment of the
lateral stress profile, sometimes also called the bending moment. The
formula is \cite{lomholt06c}
\begin{equation}
K C_0=\int_{-\infty}^\infty d z\;z\left[T^{(0)}_{x x}-\left(-p^0\right)\right]\;,
\end{equation}
and insertion of the zeroth order solution gives
\bes
K C_0=\frac{\varepsilon_0\varepsilon_1}{4}A_1^2(1-2 \kappa_1 d)-\frac{\varepsilon_0\varepsilon_3}{4}A_3^2(1-2 \kappa_3 d)\;, 
\ees
or explicitly
\begin{equation}
K C_0=\frac{\varepsilon_0 }{4}\left(\frac{\Delta \phi}{2}\right)^2 \Big( \varepsilon_1 l_1^2 (1-2 \kappa_1 d)-\varepsilon_3 l_3^2 (1-2 \kappa_3 d) \Big) \Gamma^2,\label{eq:C0}
\end{equation}
where $\Gamma= 1/[l_1 + l_3 +d/\varepsilon_2] $,
$l_1=(\varepsilon_1\kappa_1)^{-1}$ and
$l_3=(\varepsilon_3\kappa_3)^{-1}$.  Note that $C_0$ vanishes in the
symmetric case where $\kappa_1=\kappa_3$ and
$\varepsilon_1=\varepsilon_3$ as it should.

%%%%%%%%%%%%%%% Bibliography %%%%%%%%%%%%%%%%%%%%


\begin{thebibliography}{10}
\bibitem{note}
Note that the membrane thickness is not uniform along the membrane for this
  type of displacement.
\end{thebibliography}
\end{document}